\documentclass[12pt]{article}
\usepackage{amsmath, amssymb, graphicx, hyperref}
\usepackage{geometry}
\usepackage{subcaption}
\usepackage{float}
\usepackage{tikz}
\usepackage{color}
\usepackage{setspace}
\usepackage{titlesec}
\usepackage{caption}
\usepackage[numbers,sort&compress]{natbib}
\usepackage{url}
\usepackage{ragged2e}
\usepackage{mathrsfs}

\titleformat{\section}[block]{\large\bfseries}{\thesection.}{0.5em}{}
\titleformat{\subsubsection}[block]{\normalsize\bfseries}{\thesubsubsection}{0.5em}{}

\title{\textbf{A Comprehensive Review of Core-Periphery and Community Detection Paradigms}}

\author{
Imran Ansari$^{1,*}$, Pawanesh Pawanesh$^{2,*}$ \\
\small $^1$ Department of Management Studies, Indian Institute of Science, Bangalore, India \\
\small $^2$ Jindal Centre for Digital Sciences, O.P. Jindal Global University, Haryana, India \\
\small Corresponding authors: *\texttt{imran.ansari@iisc.ac.in, pawanesh@jgu.edu.in}
}

\date{}

\begin{document}

\maketitle

\begin{abstract}
Meso-scale structures, such as \textit{core-periphery (CP)} and \textit{community structure}, have attracted significant attention in modern network science. While \textit{communities} are characterized by dense intra-group and sparse inter-group connections, \textit{\textit{CP structure}s} consist of a densely interconnected core and a loosely connected periphery, where peripheral nodes are typically linked to the core. Despite growing interest, identifying \textit{\textit{CP structure}s} remains an ill-posed problem, with no universally accepted definition or standardized detection methodology. This ambiguity has led to conceptual overlaps, inconsistent evaluation metrics and slowed methodological progress. In this review, we provide a structured overview of foundational concepts, recent advances, key challenges and comparative evaluations of CP detection approaches, along with a discussion of their interplay with \textit{community structure} and applications in real-world networks.
\end{abstract}

\begin{center}
\end{center}

\newpage

\section{Introduction}
\justifying
The real world is composed of an intricate web of complex systems that pervade nearly every domain of science and society~\cite{newman2018networks}. From molecular interactions within biological cells to information transfer across technological infrastructures and the dynamics of global financial markets, complexity emerges as a fundamental characteristic of collective behavior~\cite{gulati2000strategic}. Consequently, the study of complex networks has become a unifying framework across multiple scientific disciplines, including physics, computer science, sociology, biology and economics \cite{de2005complex,costa2008complex,albert2002statistical,sun2021robustness,newman2003structure,boccaletti2006complex,lin2013complex,haznagy2015complex,mata2020complex}. These networks—ranging from social interaction graphs and transportation infrastructures to biological systems and financial markets~\cite{de2005complex,costa2008complex,albert2002statistical,sun2021robustness,newman2003structure,boccaletti2006complex,lin2013complex,haznagy2015complex,mata2020complex,spirin2003protein,tornow2003functional,chen2006detecting,girvan2002community,newman2004finding,bargigli2013finding, chan2018systemic,zhao2021community,huang2021community,pawanesh2025exploring, nie2018constructing, pawanesh2025exploiting} —exhibit rich structural patterns that profoundly influence their dynamical and functional properties. Understanding how these networks are organized has become central to revealing the principles governing their robustness, efficiency and dynamics. 

Within network science, a major research focus lies in uncovering the organizational patterns that exist between the microscopic (individual nodes and links) and macroscopic (global network) levels~\cite{tuncc2015unifying, rombach2017core}. These intermediate patterns are known as mesoscale structures. Common examples include \textit{community structure}, \textit{\textit{core-periphery (CP)} structures}, bipartite patterns, stochastic block models and multi-core configurations such as cliques, stars, cycles and motifs. Among these structures, a huge amount of literature shows that CP~\cite{borgatti2000models, holme2005core, boyd2010computing, rossa2013profiling, rombach2017core, kojaku2018core, kojaku2017finding} and \textit{community structure} \cite{newman2004finding, Newman2004,newman2006finding,blondel2008fast,raghavan2007near,lou2013detecting,zhang2017label,berahmand2018lp,shen2010covariance,xie2009detection,ansari2024identifying} have received the most sustained attention in complex network research and they are also relevant for capturing both global and local organization in networks. In recent years, a handful of surveys have looked at network \textit{communities} \cite{mohamed2019comprehensive,mata2020complex,jin2021survey,dey2022community,nooribakhsh2024community,izem2024survey} and even fewer have touched on \textit{\textit{CP structure}s}~\cite{yanchenko2023core, csermely2013structure, gallagher2021clarified}—the field still lacks a unified perspective. In this article, we bring these strands together by offering a comprehensive and comparative look at mesoscale organization. Our goal is to clarify both the theoretical and practical significance of organizations, why these structures are central to understanding how networks sustain functionality, remain resilient and enable efficient information flow. \textit{\textit{CP structure}s} characterize a division of nodes into a densely interconnected ``core'' and a sparsely connected ``periphery,''~\cite{borgatti2000models} reflecting hierarchical importance, influence, or resource centralization. Unlike \textit{communities}, the CP organization emphasizes centrality and global reach rather than local cohesiveness. \textit{communities} are particularly prominent in many real-world networks, especially those exhibiting small-world properties~\cite{watts1998collective}. They can be broadly defined as groups of nodes that are more densely connected than to the rest of the network. In literature, terms such as cluster, module, or group are often used interchangeably to describe this concept~\cite{kovacs2021inherent}. A central challenge in network \textit{communities} lies in defining exactly what constitutes a “community”. Numerous definitions and algorithms have been proposed, each emphasizing different structural features. As a result, the detected community divisions may vary even within the same network, depending on the underlying definition and analytical approach~\cite{coscia2011classification}.

Although in practice both paradigms describe mesoscale organization, they capture fundamentally different aspects of network structure~\cite{yang2018structural,pawanesh2025exploring}. \textit{Communities} highlight local modular cohesion, whereas \textit{\textit{CP structure}s} emphasize global hierarchical integration \cite{csermely2013structure}. Moreover, these patterns are not mutually exclusive; many real-world networks display a complex interplay between modular and CP organization, with certain \textit{communities} exhibiting internal cores or with global cores spanning multiple \textit{communities} \cite{gamble2016node}. The study of these paradigms is not merely of structural interest—it has profound implications in the real world. In social networks, \textit{communities} identify interest groups \cite{girvan2002community}, while cores represent influential actors or opinion leaders. In transportation systems, \textit{communities} correspond to regional connectivity clusters \cite{de2013community}, while the CP split can uncover central hubs~\cite{lee2014density} critical for network efficiency. In biological networks, \textit{communities} may correspond to functional pathways \cite{wilson2016discovery} and core nodes often represent essential biomolecules or regulators~\cite{luo2009core}. In financial systems, \textit{communities} can indicate market sectors, whereas cores can reveal systemically important institutions whose failure could trigger systemic risk~\cite{pawanesh2025exploring}. 

Over the past two decades, significant progress has been made in the development of algorithms and theoretical frameworks to detect and characterize both \textit{communities} and \textit{\textit{CP structure}s}. Community detection methods range from modularity optimization \cite{newman2004fast,clauset2004finding,wakita2007finding,blondel2008fast} and spectral techniques to statistical inference and label propagation\cite{newman2013spectral,newman2006finding,raghavan2007near,xie2009detection,zhang2017label, berahmand2018lp,shen2010covariance}. Community detection in networks has evolved rapidly in recent years, moving well beyond classical modularity-based or spectral methods to embrace machine learning, deep learning and dynamic/temporal settings. For example, the review by A Survey of Community Detection Approaches: From Statistical Modeling to Deep Learning (Jin et al., 2021) \cite{jin2021survey} provides a detailed taxonomy of methods from probabilistic graphical models to modern graph neural networks.  Meanwhile, Community detection in social networks using machine learning: a systematic mapping study (2024) \cite{nooribakhsh2024community} examines how machine-learning methods (especially unsupervised and increasingly deep-learning) are applied to community detection in social networks.  And in the more specific domain of dynamic networks, the survey A Survey On Community Identification in Dynamic Network (2024) \cite{izem2024survey} highlights how evolving network structure has pushed the field to reconsider how to track \textit{communities} over time.  These foundational reviews help set the stage for current research and highlight key challenges: how to define a “community” in heterogeneous and evolving networks, how to evaluate detected \textit{communities} reliably and how to scale methods to large real-world data. CP detection approaches range from block modeling~\cite{borgatti2000models,rombach2017core} and centrality-based methods~\cite{holme2005core,lee2014density} to geometric embedding and probabilistic generative models~\cite{rossa2013profiling, zhang2015identification}. Despite these advances, there is no universal method applicable to all contexts, as the structural signatures and functional interpretations of these patterns vary widely across domains.

The CP paradigm has transitioned from a niche complement to community detection toward a more systematically studied meso-scale feature of complex networks. In particular, \citet{yanchenko2023core} provides an up-to-date survey of statistical models, inference tasks (estimation, hypothesis testing, Bayesian frameworks) and applications across domains, consolidating CP research into a coherent statistical lens. Earlier integrative treatments, such as \citet{csermely2013structure}, characterized \textit{\textit{CP structure}s} in biological, social and infrastructural networks and contrasted them with community modules in terms of hierarchy, robustness and functional integration. More recently, the typology of CP forms has been refined by \citet{gallagher2021clarified}, which demonstrates empirically that different detection methods (e.g., k-core, two-block stochastic block models) often yield divergent partitions of core vs periphery, highlighting the need to situate one’s analysis within a clear structural taxonomy. Taken together, these works reflect a shift away from ad-hoc CP detection toward formalized frameworks, refined typologies and explicit inference agendas, setting the scene for applications in financial networks, contagion processes, mobility and comparative method studies.

This survey aims to provide a comprehensive and critical review of the methodologies, theoretical foundations and applications of CP and community detection paradigms. We systematically classify existing approaches, analyze their mathematical principles and evaluate their strengths and limitations in diverse network contexts. Furthermore, we explore the interplay between the two paradigms, highlighting emerging research directions where hybrid models and multilayer frameworks offer deeper insights into complex systems. By integrating perspectives from graph theory, statistical mechanics, network geometry and domain-specific applications, this review not only consolidates existing knowledge but also identifies open challenges and promising avenues for future research. We hope that this work will serve as a valuable reference for both theorists and practitioners seeking to advance the understanding of mesoscale structures in complex networks. Complex networks often exhibit structures beyond local node-level statistics or global degree distributions. These meso-scale structures play a crucial role in understanding the functional, structural and dynamical aspects of networks. Among them, \textit{communities} and \textit{\textit{CP structure}s} are the most studied. This review aims to provide a comprehensive discussion of both paradigms, their algorithms, theoretical foundations and applications across disciplines.

The remainder of this paper is organized into six sections to provide a comprehensive understanding of the CP and \textit{community structure} and their applications in complex systems. \textit{Section 2} introduces the \textit{CP structure} in networks, highlighting the principal methodologies developed for its detection, the diverse domains in which it has been applied and its growing significance in understanding hierarchical organization and systemic stability across real-world systems. \textit{Section 3} presents a detailed discussion of the \textit{community structure} covering the foundational approaches to community detection, the role of community information in network analysis and its practical implications in the financial, biological and social domains. The section emphasizes how modular structures facilitate a deeper understanding of functional organization. \textit{Section 4} provides a comparative examination of the CP and \textit{community structure} with particular attention to their sectoral representations and the distinct but complementary insights they offer for portfolio optimization.
Furthermore, this section explores the manifestation of both mesoscale structures within the framework of hyperbolic network models, offering a geometric perspective on their coexistence. \textit{Section 5} investigates the similarities and dissimilarities between these mesoscale structures, highlighting their topological, functional and interpretational distinctions as well as the interplay that underpins the collective dynamics of complex systems. Finally, \textit{Section 6} discusses open challenges and future research directions, outlining unresolved questions and emerging opportunities in the exploration of CP and \textit{community structure}, particularly their integration into dynamic, multilayer and higher-order network frameworks with real-world applications.

\section{\textit{CP structure} in Networks}

This section provides a comprehensive overview of the \textit{CP structure} in complex networks, detailing its detection methodologies, diverse real-world applications and its fundamental significance in revealing hierarchical organization and functional dynamics within networks. The subsections include: (i). Methods for detecting the \textit{CP structure} in networks, (ii). Applications of \textit{CP structure} in various fields and (iii). Significance of \textit{CP structure} in networks. 

\subsection{Methods for detecting the \textit{CP structure} in networks}

The \textit{CP structure} provides a powerful framework for understanding the heterogeneous organization of complex networks. In such a structure, the densely connected core of nodes coexists with a sparsely connected periphery, whose members are loosely connected but are often linked to the core. This pattern has been observed in various systems, including social and communication networks, as well as financial and transportation systems. One of the earliest and most influential formalizations of this idea was proposed by Borgatti and Everett in 1999, who developed a block model-based approach to identify and quantify the organization of the CP in networks.

One of the first formal definitions of a \textit{CP structure}, proposed by Borgatti \textit{et al.}~\cite{borgatti2000models}, considers a network represented as an undirected, unweighted \(G = (V, E)\) with \(N\) vertices and \(M\) edges and no self-loops or multiple edges. Let \(A = (a_{ij})\) denote the adjacency matrix, where \(a_{ij} = 1\) if vertices \(i\) and \(j\) are adjacent and \(a_{ij} = 0\) otherwise. The network \(G\) is said to exhibit a \textit{CP structure} if there exists a subset of \textit{core vertices} \(\mathscr{K} \subset V\) and a subset of \textit{peripheral vertices} \(\mathcal{P} \subset V \setminus \mathscr{K}\) such that:
\begin{itemize}
    \item[i.] For all \(i, j \in \mathscr{K}\): \( a_{ij} = 1 \).
    \item[ii.] For all \(i, j \in \mathcal{P}\): \( a_{ij} = 0 \).
    \item[iii.] For each \(i \in \mathscr{K}\), there exists at least one \(j \in \mathcal{P}\) with \( a_{ij} = 1 \); and for each \(j \in \mathcal{P}\), there exists at least one \(i \in \mathscr{K}\) with \( a_{ij} = 1 \).
\end{itemize}

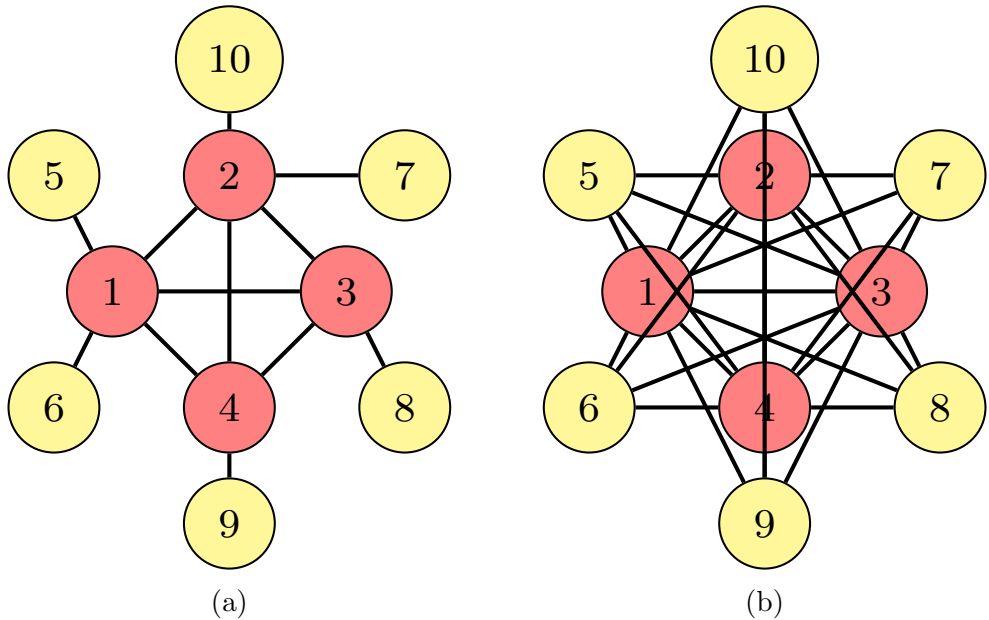
\begin{figure}[http]
    \centering
    \begin{subfigure}[b]{0.40\textwidth}
        \centering
        \resizebox{\linewidth}{!}{
            \begin{tikzpicture}[scale=0.8, every node/.style={draw, circle, minimum size=0.2cm, font=\scriptsize}]
                \node[fill=red!50] (C1) at (0, 1.5) {1};
                \node[fill=red!50] (C2) at (1, 2.5) {2};
                \node[fill=red!50] (C3) at (2, 1.5) {3};
                \node[fill=red!50] (C4) at (1, 0.5) {4};
                
                \node[fill=yellow!50] (P1) at (-0.5, 2.5) {5};
                \node[fill=yellow!50] (P2) at (-0.5, 0.5) {6};
                \node[fill=yellow!50] (P3) at (2.5, 2.5) {7};
                \node[fill=yellow!50] (P4) at (2.5, 0.5) {8};
                \node[fill=yellow!50] (P5) at (1, -0.5) {9};
                \node[fill=yellow!50] (P6) at (1, 3.5) {10};

                \draw[thick] (C1) -- (C2);
                \draw[thick] (C1) -- (C3);
                \draw[thick] (C1) -- (C4);
                \draw[thick] (C2) -- (C3);
                \draw[thick] (C2) -- (C4);
                \draw[thick] (C3) -- (C4);

                \draw[thick] (C1) -- (P1);
                \draw[thick] (C1) -- (P2);
                \draw[thick] (C2) -- (P3);
                \draw[thick] (C3) -- (P4);
                \draw[thick] (C4) -- (P5);
                \draw[thick] (C2) -- (P6);
            \end{tikzpicture}
        }
        \caption{}
        \label{fig:realistic_network}
    \end{subfigure}
    \hspace{0.4cm}
    \begin{subfigure}[b]{0.40\textwidth}
        \centering
        \resizebox{\linewidth}{!}{
            \begin{tikzpicture}[scale=0.8, every node/.style={draw, circle, minimum size=0.2cm, font=\scriptsize}]
                \node[fill=red!50] (IC1) at (0, 1.5) {1};
                \node[fill=red!50] (IC2) at (1, 2.5) {2};
                \node[fill=red!50] (IC3) at (2, 1.5) {3};
                \node[fill=red!50] (IC4) at (1, 0.5) {4};

                \node[fill=yellow!50] (IP1) at (-0.5, 2.5) {5};
                \node[fill=yellow!50] (IP2) at (-0.5, 0.5) {6};
                \node[fill=yellow!50] (IP3) at (2.5, 2.5) {7};
                \node[fill=yellow!50] (IP4) at (2.5, 0.5) {8};
                \node[fill=yellow!50] (IP5) at (1, -0.5) {9};
                \node[fill=yellow!50] (IP6) at (1, 3.5) {10};

                \draw[thick] (IC1) -- (IC2);
                \draw[thick] (IC1) -- (IC3);
                \draw[thick] (IC1) -- (IC4);
                \draw[thick] (IC2) -- (IC3);
                \draw[thick] (IC2) -- (IC4);
                \draw[thick] (IC3) -- (IC4);

                \foreach \core in {IC1, IC2, IC3, IC4}
                    \foreach \peri in {IP1, IP2, IP3, IP4, IP5, IP6}
                        \draw[thick] (\core) -- (\peri);
            \end{tikzpicture}
        }
        \caption{}
        \label{fig:ideal_network}
    \end{subfigure}
    \caption{Illustration of a network with (a) a \textit{CP structure} and (b) an ideal \textit{CP structure}}
    \label{fig:core_periphery_structures}
\end{figure}

Figure~\ref{fig:core_periphery_structures} displays an example of a \textit{CP structure} and an ideal \textit{CP structure}. It has ten vertices, of which four are in the core. In the \textit{CP structure}, the core vertices (light red) are adjacent to other core vertices and some periphery vertices (light yellow), while the periphery vertices are not adjacent. However, in the ideal \textit{CP structure}, all core vertices are adjacent and all periphery vertices are not adjacent.

Later, numerous methods were developed, expanding the concept to weighted, directed and temporal networks and offering alternative formulations. Detecting \textit{\textit{CP structure}s} is crucial for understanding the functional organization of complex networks across diverse domains. A variety of methods have been developed for this purpose, including statistical inference~\cite{zhang2015identification,peixoto2019bayesian,kojaku2017finding}, spectral analysis~\cite{rombach2017core,cucuringu2016detection}, diffusion processes~\cite{rossa2013profiling}, motif-based measures~\cite{ma2018detection}, geodesic patterns~\cite{cucuringu2016detection} and model-based approaches~\cite{rombach2017core}. While these methods offer a rich diversity of perspectives, they are often built on differing---and sometimes conflicting---assumptions about how core and peripheral vertices connect and how such structures manifest in networks~\cite{malliaros2020core}. This variation means that applying a given method without understanding its underlying assumptions can lead to inconsistent interpretations. Consequently, researchers and practitioners must choose the method that aligns with their conceptualization of the CP patterns. A clear typology and robust statistical tools for identifying distinct \textit{\textit{CP structure}s} can ensure methodological consistency and enable reliable inferences about structural and dynamical properties of networks. In the following subsection, we present the key models, beginning with the fundamental work of Borgatti \textit{et al.} on detecting \textit{\textit{CP structure}s} in networks.


In~\cite{borgatti2000models}, Borgatti \textit{et al.} proposed two types of CP models, namely, the \emph{discrete} and \emph{continuous} versions.
In the discrete model, each vertex \( i \) is assigned a label \( c_i \) according to the rule: \( c_i = 1 \) if vertex \( i \) belongs to the core set \( \mathscr{K} \) and \( c_i = 0 \) if vertex \( i \) belongs to the periphery set \( \mathcal{P} \).
The sets \( \mathscr{K} \) and \( \mathcal{P} \) are determined by maximizing the function

\begin{equation}
Q_{\text{cp}}(c_1,\ldots,c_N) = \sum_{i=1}^{N} \sum_{j=1}^{N} a_{ij} c_{ij},
\label{eq:cp1_quality}
\end{equation}

over all \( c_1, \ldots, c_N \), where \( a_{ij} \) denotes the elements of the adjacency matrix.
Here, \( c_{ij} \) is set to 1 if \( i \) or \( j \) belong to \( \mathscr{K} \) and set to 0 if both belong to \( \mathcal{P} \).

In the continuous model, the idea of assigning a \textit{coreness} value to each vertex is introduced. Specifically, a quantity \( c_i \in [0,1] \) is assigned to each vertex \( i \), representing its degree of belonging to the core. These values are computed by maximizing  
\begin{equation}
Q_{\text{cp}}(c_1, \ldots, c_N) = \sum_{i=1}^{N} \sum_{j=1}^{N} a_{ij} c_i c_j.
\label{eq:cp2_quality}
\end{equation}

In this formulation, \textit{coreness} value provides a quantitative measure of the affiliation of then of the vertices with the core or the periphery. A high \textit{coreness} value indicates that the vertex is more likely to be part of the core, whereas a low value suggests an association with the periphery. In particular, a value of \( c_i \) close to \( 1 \) implies that vertex \( i \) is more likely to be a \emph{pure core} vertex, while a value close to \( 0 \) implies that it is more likely to be a \emph{pure peripheral} vertex.


In~\cite{holme2005core}, Holme introduced a \textit{CP coefficient}, denoted as \( c^{cp}(G) \), to quantify the extent to which a given network \( G \) exhibits a \textit{CP structure}:

\begin{equation}
c^{cp}(G) = \frac{C^{C}(V_{\text{core}}(G))}{C^{C}(V(G))} - \left\langle \frac{C^{C}(V_{\text{core}}(G'))}{C^{C}(V(G'))} \right\rangle_{G' \in \mathcal{G}(G)},
\end{equation}

\noindent where \( V(G) \) is the set of vertices in the graph \( G \), \( V_{\text{core}}(G) \) is the \textit{\( k \)-core}\index{\( k \)-core} (maximal subgraph in which each vertex has degree at least $k$ within the subgraph) that maximizes closeness centrality and \( \mathcal{G}(G) \) represents an ensemble of random graphs with the same degree distribution as \( G \). The \textit{closeness centrality of a subgraph} \( U \) is defined as:

\begin{equation}
C^{C}(U) = \left\langle \left\langle P(i, j) \right\rangle_{j \in V \setminus \{i\}} \right\rangle_{i \in U}
= \frac{1}{|U|} \sum_{i \in U} \left( \frac{1}{|V|-1} \sum_{j \in V \setminus \{i\}} P(i, j) \right)
\end{equation}

\noindent where \( P(i,j) \) is the distance between the vertices \( i \) and \( j \).


In~\cite{da2008centrality}, Da Silva \textit{et al.} introduced \textit{network capacity}, a measure of overall connectivity in a network. It is defined as the sum of the inverse of the shortest path lengths between all connected pairs of vertices in the network. The \textit{network capacity} is defined and denoted as follows:
\begin{equation}
K = \sum_{l=1}^{M} P_l^{-1}
\end{equation}

Where \( M \) and \( P_l \) are the total number of pairs of vertices and the length of the shortest path between the \( l \)-th pair of vertices in the network. A higher value of \( K \) suggests that the network is more densely connected, which means that the vertices are, on average, closer to each other.


In~\cite{boyd2010computing}, Boyd \textit{et al.} aim to find a \textit{coreness or MINRES vector} \( \mathbf{c} \) such that the adjacency matrix \( \mathbf{A} \) is approximated by \( \mathbf{c}\mathbf{c}^T \). This approximation minimizes the sum of squared differences off the diagonal. Therefore, they find a vector \( \mathbf{c} \) by minimizing
\begin{equation}
Q_{\text{cp}}(c_1, c_2, \ldots, c_N) = \sum_{i} \sum_{j \neq i} \left( a_{ij} - c_i c_j \right)^2.
\end{equation}

After partially differentiating $Q_{\text{cp}}$ with respect to each element of \( \mathbf{c} \), it leads to an iterative process to calculate the coreness vector:
\begin{equation}
c_i = \frac{\sum_{j \neq i} a_{ij} c_j}{\sum_{j \neq i} c_j^2}.
\end{equation}

In~\cite{rossa2013profiling}, Rossa \textit{et al.} proposed an intuitive approach to identify the core and periphery by modeling a random-walker Markov chain process, which avoids arbitrary partitions of the network. The vertices of the weighted network represent the set of states of the Markov chain and \([p_{ij}]\) is the matrix of transition probabilities from vertex \( i \) to \( j \), defined in terms of the weighted adjacency matrix \([a_{ij}]\) as
\begin{equation}
\label{eq:normalized_probability}
p_{ij} = \frac{a_{ij}}{\sum_h a_{ih}}.
\end{equation}

The objective is to find the largest subset \(S\) of vertices such that if a random walker starts from any vertex in \(S\), there is a high probability that it escapes from \(S\) in the next step; in other words, there is only a very small probability of remaining within \(S\). This implies that the connectivity among the vertices in \(S\) is either nonexistent or very weak and thus, \(S\) can be interpreted as the periphery of the network. Mathematically, the probability that a random walker remains within the subset \(S\) is defined as
\begin{equation}
\label{eq:community_centrality}
\phi_S = \frac{\sum_{i,j \in S} \pi_i p_{ij}}{\sum_{i \in S} \pi_i},
\end{equation}
where \(\pi_i > 0\) denotes the asymptotic probability of being at vertex \(i\). Owing to the irreducibility of the Markov chain, this expression simplifies to the following condition:
\[
\pi_j = \sum_{i \in V(G)} \pi_i p_{ij} = \sum_{i \in V(G)} \pi_i \frac{a_{ij}}{\sum_{h \in V(G)} a_{ih}}.
\]
For additional details, readers are referred to~\cite{meyer2023matrix}. Through a simple algebraic manipulation, one obtains
\begin{equation}
\label{eq:node_centrality}
\pi_i = \frac{C^{w}_{DC}(i)}{\sum_{j \in V(G)} C^{w}_{DC}(j)},
\end{equation}
where \(C^{w}_{DC}(i) = \sum_h a_{ih}\) represents the strength or weighted degree of vertex \(i\). Consequently, \(\phi_S\) can be expressed in a simplified form, making it straightforward to compute:
\begin{equation}
\label{eq:community_connectivity}
\phi_S = \frac{\sum_{i,j \in S} a_{ij}}{\sum_{i \in S, j \in V(G)} a_{ij}}.
\end{equation}

The set \(S\) is constructed iteratively as follows: start with a vertex of minimum weighted degree and denote the initial set containing this vertex as \(S_1\); without loss of generality, let \(S_1 = \{1\}\), for which we define \(\phi_1 := \phi_{S_1} = 0\). In the next step, consider all subsets \(S_2^{(j)} := S_1 \cup \{j\}\) for \(2 \leq j \leq N\) and compute the persistence probabilities \(\phi_{S_2^{(j)}}\); let \(\phi_{S_2^{(k)}}\) be the minimum among them, then define \(S_2 := S_2^{(k)}\) and \(\phi_2 := \phi_{S_2^{(k)}} = \phi_{S_2}\), assigning this value as the \textit{coreness} of vertex \(k\), so that \(\phi_1 \leq \phi_2\). This procedure is repeated iteratively: at each step \(m\), construct \(S_m\) by adding a new vertex that minimizes the persistence probability among all candidates and assign \(\phi_m\) as its \textit{coreness}, ensuring \(\phi_1 \leq \phi_2 \leq \cdots \leq \phi_m\). Continuing until all vertices are included, we obtain the ``CP profile" of the network: \(\phi_1 \leq \phi_2 \leq \cdots \leq \phi_N\).


In Ref.~\cite{rombach2017core}, Rombach \textit{et al.} framed core--periphery detection as an optimization problem in which each vertex is assigned a \textit{coreness} value. For a weighted and undirected network with adjacency matrix \(A=(a_{ij})\), they considered a shuffled vector of coreness scores \((c_1, c_2, \ldots, c_N)\) and searched for the permutation that maximizes a quality function depending on parameters \(\alpha\) and \(\beta\):
\[
Q_{\text{cp}}(\alpha, \beta)
= \sum_{i=1}^{N} \sum_{j=1}^{N} a_{ij}\, c_i(\alpha, \beta)\, c_j(\alpha, \beta).
\]

The starting \textit{coreness} profile \(c_i^*(\alpha,\beta)\) is constructed using two parameters $\alpha$ and $\beta$ in \([0,1]\). Here, \(\beta\) controls the proportion of vertices treated as \textit{core} and \(\alpha\) regulates how sharply \textit{coreness} values separate \textit{core} from \textit{periphery}. The initial values are defined as
\[
c_i^*(\alpha, \beta) =
\begin{cases}
\dfrac{i(1 - \alpha)}{2\lfloor \beta N \rfloor}, & i \le \lfloor \beta N \rfloor, \\[6pt]
\dfrac{(i - \lfloor \beta N \rfloor)(1 - \alpha)}{2(N - \lfloor \beta N \rfloor)} + \dfrac{1 + \alpha}{2}, & i > \lfloor \beta N \rfloor.
\end{cases}
\]

By permuting these values over all vertices, algorithm identifies the ordering \((c_i)\) that maximizes \(Q_{\text{cp}}(\alpha,\beta)\). Repeating this process for many choices of \((\alpha,\beta)\) allows each vertex to accumulate evidence of its involvement in the core. The aggregated core score of vertex \(i\) is then
\[
CS(i) = Z \sum_{\alpha,\beta} c_i(\alpha,\beta)\, Q_{\text{cp}}(\alpha,\beta),
\]
where \(Z\) is chosen to ensure that the maximum core score will be 1. Equivalently,
\[
CS(i)
=
\frac{\displaystyle \sum_{\alpha,\beta} c_i(\alpha,\beta)\, Q_{\text{cp}}(\alpha,\beta)}
{\displaystyle \max_{1 \le j \le N}
\sum_{\alpha,\beta} c_j(\alpha,\beta)\, Q_{\text{cp}}(\alpha,\beta)}.
\]


In the work of Zhang \textit{et al.}~\cite{zhang2015identification}, authors imagine the task of uncovering a hidden \textit{CP structure} in a network as a detective-like process. The observed network is undirected and unweighted, represented by an adjacency matrix \(A \in \{0,1\}^{n \times n}\). Beneath this observed structure, they assume a latent generative mechanism based on a two-group Stochastic Block Model (SBM). Each vertex \(i\) secretly belongs to one of two hidden classes: the \textit{core} (\(r=1\)) or the \textit{periphery} (\(r=2\)) and the algorithm’s objective is to infer these unknown memberships. To begin, the model assigns prior probabilities to the two groups, denoted as \(\boldsymbol{\gamma} = [\gamma_1, \gamma_2]\), where \(\gamma_r = \mathbb{P}(z_i = r)\) and \(\sum_{r=1}^2 \gamma_r = 1\). The interaction pattern between vertices is governed by a connectivity matrix \(\mathbf{P} = (p_{rs}) \in [0,1]^{2 \times 2}\), where the \textit{CP structure} is encoded in the constraint
\[
p_{11} > p_{12} > p_{22}.
\]
The model cannot observe group labels directly, so it relies on ``responsibility messages’’ \(\eta_{ij}^{(r)}\), which represent how strongly vertex \(i\) signals to vertex \(j\) that it belongs to group \(r\). These messages are updated within an Expectation--Maximization (EM) framework, where Belief Propagation (BP) forms the core of the E-step. The BP process begins with randomly initialized and normalized messages satisfying \(\sum_{r=1}^2 \eta_{ij}^{(r)} = 1\). They are updated iteratively according to
\[
\eta_{ij}^{(r)} \propto \gamma_r 
\prod_{\substack{k \neq i,j \\ A_{ik}=1}}
\left( \sum_{s=1}^2 \eta_{ki}^{(s)} p_{rs} \right),
\]
followed by a damping adjustment using a factor \(\alpha \in (0,1)\):
\[
\eta_{ij}^{(r)} \leftarrow 
\alpha\, \eta_{ij}^{(r)} 
+ (1 - \alpha)\, \eta_{ij,\text{new}}^{(r)}.
\]
As these messages stabilize, they allow the algorithm to estimate the marginal probability that vertex \(i\) belongs to group \(r\). These vertex-level marginals are computed as
\[
q_i^r \propto \gamma_r 
\prod_{\substack{k \neq i \\ A_{ik}=1}}
\left( \sum_{s=1}^2 \eta_{ki}^{(s)} p_{rs} \right),
\]
and normalized such that \(\sum_{r=1}^2 q_i^r = 1\). With updated beliefs about the latent group assignments, the algorithm refines the model parameters. The prior probabilities are updated using
\[
\gamma_r = \frac{1}{n} \sum_{i=1}^n q_i^r,
\]
while the edge probabilities in \(\mathbf{P}\) are recalibrate using expected counts of edges:
\[
p_{rs} = 
\frac{
\sum\limits_{i<j} A_{ij} q_i^r q_j^s
}{
\sum\limits_{i<j} q_i^r q_j^s
}.
\]

Once convergence is achieved, the model evaluates how well the inferred \textit{CP structure} explains the observed network. The log-likelihood under the SBM is computed by
\[
\mathcal{L}_{\text{CP}} = 
\sum_{i<j} \sum_{r,s} q_i^r q_j^s 
\left[ A_{ij} \log p_{rs} 
+ (1 - A_{ij}) \log (1 - p_{rs}) \right]
+
\sum_{i=1}^n \sum_{r=1}^2 q_i^r \log \gamma_r.
\]

To determine whether the detected CP pattern is meaningful or merely a consequence of degree heterogeneity, the model compares this likelihood against a configuration-model null hypothesis. Under this null model, the probability of an edge between vertices \(i\) and \(j\) is determined solely by their degrees:
\[
p^{\text{null}}_{ij} = \frac{k_i k_j}{2m}, \qquad
m = \frac{1}{2} \sum_{i,j} A_{ij}, \qquad
k_i = \sum_j A_{ij}.
\]
The corresponding log-likelihood is given by
\[
\mathcal{L}_{\text{null}} =
\sum_{i<j}
\left[
A_{ij} \log p^{\text{null}}_{ij}
+ (1 - A_{ij}) \log \left(1 - p^{\text{null}}_{ij}\right)
\right].
\]

Therefore, the strength of the CP organization can be quantified through the quality score
\[
Q_{\text{cp}} = 
\frac{\mathcal{L}_{\text{CP}} - \mathcal{L}_{\text{null}}}
{|\mathcal{L}_{\text{null}}|},
\]
where a larger value of \(Q_{\text{cp}}\) indicates a more pronounced and statistically meaningful \textit{CP structure} beyond what is expected from degree-based connectivity alone.


In~\cite {jia2019random}, Jia \textit{et al.} adopted a probabilistic approach to detect the \textit{CP structure} in undirected and unweighted networks. Let $c = [c_1, c_2, \dots, c_N]^T$ denote the vector of core scores. Each node $i \in V$ is assigned a \emph{core score} $c_i$, 
nodes with larger $c_i$ values are more ``core-like,'' while nodes with lower values of $c_i$ are more peripheral. The probability $\rho_{ij}$ of an edge that exists between nodes $i$ and $j$ depends on their core scores $c_i, c_j$ and their spatial distance $d_{ij}$ and is defined as
\begin{equation}
    \rho_{ij} = \frac{e^{c_i + c_j}}{e^{c_i + c_j} + d_{ij}^{\epsilon}},
\end{equation}
where $\epsilon > 0$ is a hyperparameter controlling the influence of distance on connectivity. 
This ensures that nodes with high core scores are more likely to connect, while spatially distant nodes have a reduced likelihood of connectivity.

Given an observed adjacency matrix $A$, the probability distribution over its entries can be written as
\begin{equation}
    P([A]_{ij}) =
    \left( \frac{e^{c_i + c_j}}{e^{c_i + c_j} + d_{ij}^{\epsilon}} \right)^{[A]_{ij}}
    \left( \frac{d_{ij}^{\epsilon}}{e^{c_i + c_j} + d_{ij}^{\epsilon}} \right)^{(1 - [A]_{ij})}.
\end{equation}

The overall log-likelihood function of the observed network is then expressed as
\begin{equation}
    \mathcal{L}(c) =
    \sum_{(i,j) \in E} \log \left( \frac{e^{c_i + c_j}}{e^{c_i + c_j} + d_{ij}^{\epsilon}} \right)
    + \sum_{(i,j) \in \bar{E}} \log \left( \frac{d_{ij}^{\epsilon}}{e^{c_i + c_j} + d_{ij}^{\epsilon}} \right),
\end{equation}
where $\bar{E}$ denotes the complement of the set of edges in the graph. 

Maximizing $\mathcal{L}(c)$ with respect to the core score vector $c$ yields the estimated \textit{coreness} of nodes. 
The resulting structure naturally separates the densely interconnected core nodes from the sparsely connected peripheral nodes.


In~\cite{kojaku2018core}, Kojaku and Masuda introduced an algorithm for detecting multiple core--periphery (CP) pairs in a network while accounting for degree heterogeneity through the configuration model as a null model. They assume that the network consists of \( C \) non-overlapping CP pairs, each comprising one core block and one periphery block. Each CP pair is characterized by the following structural properties:  
(1) dense intra-core connectivity,  
(2) dense core--periphery connectivity,  
(3) sparse intra-periphery connectivity and  
(4) sparse interactions between different CP pairs.

Their study~\cite{kojaku2018core} aim is to identify \( c_i \) (\(1\leq c_i \leq C\)) and \( x_i \), where \( c_i \) is the index of the CP pair to which vertex \( i \) belongs and \( x_i = 1 \) or \( x_i = 0 \) if the corresponding vertex is a core or peripheral vertex, respectively, by maximizing \( Q_{\text{cp}} \), which is defined as

\begin{equation}
Q_{\text{cp}} = \frac{1}{2M} \sum_{i=1}^N \sum_{j=1}^N 
\left( A_{ij} - \frac{d_i d_j}{2M} \right) 
(x_i + x_j - x_i x_j) \delta(c_i, c_j).
\end{equation}

Here, \( \delta(c_i, c_j) \) is the Kronecker delta, \( 2M \) is the total number of edges and \( d_i \) and \( d_j \) are the degrees of vertices \( i \) and \( j \), respectively. If all vertices are core vertices (i.e. \( x_i = 1 \) for all \( i \)), then \( Q_{\text{cp}} \) becomes equivalent to the modularity of the network \cite{newman2004finding, newman2006finding}. The rationale for defining the above expression is to maximize the similarity between the ground truth adjacency matrix and the idealized \textit{CP structure} that satisfies the properties (1)–(4) mentioned above.

\subsection{Application of \textit{CP structure} in various fields}

In network science, CP network structures have gained significant attention in recent years~\cite{borgatti2000models, boyd2010computing, kojaku2018core, rombach2017core}. Algorithmic detection of these \textit{\textit{CP structure}s} enables the discovery of network characteristics that are not readily observable at the local level of nodes and edges or through global summary statistics. In a \textit{CP structure}, the core nodes are densely interconnected and connected to some periphery nodes, while the periphery nodes are not directly connected \cite{borgatti2000models, holme2005core, rossa2013profiling}. This configuration reflects a hierarchical and functional organization within the network. This CP pattern has helped explain a wide range of networked phenomena, including online amplification~\cite{barbera2015critical}, cognitive learning processes~\cite{bassett2013task}, organization of technological infrastructure~\cite{alvarez2005k,carmi2007model} and critical disease spreading conduits~\cite{kitsak2010identification}. It applies seamlessly across domains by providing a succinct mesoscale description of a network’s organization around its core. By decomposing a network into core and peripheral nodes, the \textit{CP structure} distinguishes central processes from those at the margins, enabling a more precise classification of nodes’ functional and dynamical roles concerning their structural position. The analytic generality of this approach, together with the relative ubiquity of CP organization among networks, makes it an indispensable methodological concept in the network science toolkit.

During the past few decades, the \textit{CP structure} has found applications in various domains, including social networks~\cite{borgatti2000models, Boyd2010,rossa2013profiling}, protein--protein interaction (PPI) networks~\cite{rossa2013profiling, Yang2014overlapping}, financial networks~\cite{craig2010interbank,van2016formation}, transportation networks~\cite{rombach2017core,rossa2013profiling}, brain networks~\cite{bassett2013task}, metabolic networks~\cite{da2008centrality}, international trade networks~\cite{boyd2010computing,rossa2013profiling,ma2015rich}, coauthorship networks~\cite{rombach2017core} and neural networks~\cite{rossa2013profiling, Tunc2015}. For instance, in a coauthorship network of researchers, leading researchers often publish papers with other leading researchers, forming the core, whereas other researchers tend to collaborate primarily with specific leading researchers---often those within the same research group---thus forming the periphery. A similar pattern emerges in world trade networks, where economically strong countries tend to trade extensively with other strong countries, forming the core. In contrast, economically weaker countries primarily trade with strong countries, forming the periphery. Numerous methods have been developed to detect and analyze \textit{\textit{CP structure}s}, highlighting their significance in understanding network dynamics and information flow across diverse fields~\cite{borgatti2000models, Boyd2010,rossa2013profiling,kojaku2018core,rombach2017core}.

A literature survey shows that notable work has been done in analyzing the \textit{CP structure} in financial networks. Works such as Chang and Zhang~\cite{chang2015endogenous}, Wang~\cite{wang2016core}, Bedayo \textit{et al.}~\cite{bedayo2016bargaining}, Castiglionesi and Navarro~\cite{castiglionesi2020efficient} and Farboodi~\cite{farboodi2023intermediation} emphasize the role of heterogeneity between agents, which becomes pivotal for the emergence of \textit{\textit{CP structure}s} in financial networks. For example, Farboodi~\cite{farboodi2023intermediation} shows that heterogeneity in investment opportunities leads to a competitive dynamic where investment banks, offering superior intermediation rates, naturally form the core. Similarly, Bedayo \textit{et al.}~\cite{bedayo2016bargaining} highlight that heterogeneity arises from agents' varying levels of time discounting, where impatient agents centralize to form the core through sequential and bilateral bargaining in intermediation. Castiglionesi and Navarro~\cite{castiglionesi2020efficient} focus on endogenous heterogeneity emerging from risk preferences, where some banks focus on safe investments while others pursue risky projects. A \textit{CP structure} arises as safe banks freely link among themselves, while connections to risky banks are more selective, driven by the need for liquidity shock insurance. Craig \textit{et al.}~\cite{craig2010interbank}, Lee \textit{et al.}~\cite{lee2014density} and In't Veld \textit{et al.}~\cite{van2016formation} highlight significant work on analyzing \textit{\textit{CP structure}s} in financial networks. For example, Lee \textit{et al.}~\cite{lee2014density} examined a combination of ETFs and their constituent stocks. A CP analysis of a network based on correlation, using the Rombach model~\cite{rombach2017core}, reveals that ETFs tend to occupy the core. 

\subsection{Significance of \textit{CP structure} of Networks}

 Many social, physical and biological systems can be represented as networks, where the vertices correspond to individual entities and the edges denote pairwise relationships between them~\cite{yang2013networks}. Among various structural properties, empirical networks often exhibit a \textit{CP structure}. Many algorithms have been proposed to identify the \textit{CP structure} in networks~\cite{borgatti2000models,boyd2010computing,rossa2013profiling,rombach2017core,holme2005core}. 
However, a less frequently addressed but fundamental question concerns the \emph{significance} of such structures. 
Do all networks necessarily exhibit a CP organization? Clearly, without rigorous statistical validation, one could mistakenly infer the presence of \textit{CP structure} in every network, leading to misleading conclusions with important practical implications. 
Thus, the development of robust statistical methods to assess the significance of \textit{CP structure} is essential in network science. 
In this section, we provide a detailed discussion on approaches for evaluating the statistical significance of \textit{\textit{CP structure}s}.

One of the earliest discussions on the statistical testing of \textit{\textit{CP structure}s} was provided by Borgatti and Everett~\cite{borgatti2000models}. 
They argued that a network can be said to exhibit a core--periphery organization if there is a strong correlation between the idealized core--periphery matrix and the ground-truth adjacency matrix (after ordering nodes according to their core scores). 
This assessment is carried out using the Quadratic Assignment Procedure (QAP) test, which evaluates whether the observed correlation is significantly higher than expected by chance. 
Specifically, Borgatti and Everett used Eq.~\ref{eq:cp1_quality} as the basis for this statistical validation. 
In the QAP framework, the correlation between the ideal CP matrix and the observed adjacency matrix serves as the test statistic. If this correlation is low and the QAP permutation test yields a non-significant result (commonly $p > 0.1$), one concludes that there is no evidence supporting the presence of a statistically significant \textit{CP structure} in the network.

Boyd \textit{et al.}~\cite{boyd2006computing} assessed the statistical significance of the detected \textit{CP structure} using a permutation-based approach. The quality measure \( Q \), originally proposed by Borgatti and Everett~\cite{borgatti2000models}, was compared with its distribution under a suitable null model obtained by randomizing the network while preserving the total number of edges. The \( p \)-value was then estimated as the proportion of randomized networks yielding a quality value greater than or equal to the observed \( Q \).

In Ref.~\cite{rossa2013profiling}, Rossa \textit{et al.} introduced a quantitative measure of the strength of a core--periphery organization, termed \emph{core--periphery centralization} (cp-centralization), defined as
\begin{equation}
C = 1 - \frac{2}{n - 2} \sum_{k=1}^{n-1} \phi_k,
\end{equation}
where \( (\phi_1, \ldots, \phi_n) \) represents the core--periphery profile~\cite{rossa2013profiling} of the network. The value of \( C \) lies between 0 and 1, with higher values indicating a stronger deviation from a complete graph toward an ideal \textit{CP structure}.  

To employ the statistical significance of the observed cp-centralization \( C \), the authors~\cite{rossa2013profiling} compared it against values obtained from an ensemble of randomized networks that preserve the degree distribution of the original network~\cite{milo2003uniform, zlatic2009rich}. For each realization, the cp-centralization \( C_{\mathrm{rand}} \) was computed and the significance of the empirical value \( C \) was quantified using the \( z \)-score:
\begin{equation}
z = \frac{C - \langle C_{\mathrm{rand}} \rangle}{\sigma(C_{\mathrm{rand}})},
\end{equation}
where \(\langle C_{\mathrm{rand}} \rangle\) and \(\sigma(C_{\mathrm{rand}})\) denote the mean and standard deviation of cp-centralization across the randomized ensemble. A large positive \( z \)-score indicates that the observed structure is unlikely to arise from random organization, confirming the statistical robustness of the detected core--periphery configuration.

\section{\textit{Community Structure} in Networks}

\justifying 
In this section, we present a clear and concise description of what we mean by a community in a network, followed by an introduction to modularity, one of the most widely used measures for evaluating the quality of a community partition. We then outline the mathematical ideas that form the foundation of several established community detection algorithms.

\begin{figure}[h]
    \centering
    \includegraphics[width=0.5\linewidth]{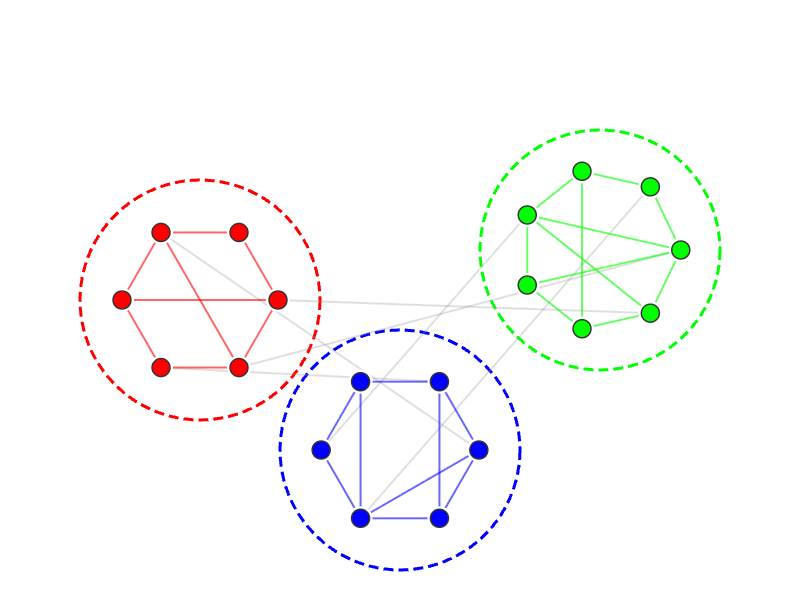}
    \caption{A simple illustration of a network divided into \textit{communities}. Nodes within the same group share denser connections than those across groups.}
    \label{fig:community}
\end{figure}

\textit{Communities}---often referred to as groups, clusters, or modules---represent regions within a complex network where vertices are more densely connected to each other than to the rest of the network (see Figure~\ref{fig:community}). Although \textit{communities} arise naturally in many real-world systems, there is still no single, universally accepted formal definition. Over the years, numerous approaches have been proposed to identify such structures, reflecting the richness and diversity of the problem~\cite{fortunato2010community, fortunato2016community, cherifi2019community}. A natural starting point is modularity~\cite{newman2004finding}, a metric designed to quantify how well a network is partitioned into internally cohesive groups. Modularity compares the density of edges inside a proposed community with the density expected under a suitable null model, thereby capturing how much the observed structure deviates from randomness.

To formalize this idea, let $c_i$ denote the community label assigned to vertex $i$, taking integer values in $\{1,\ldots,n_c\}$, where $n_c$ is the total number of \textit{communities}. Modularity is built upon measuring the difference between the actual number of edges within a community and the expected number of such edges under the null model. For vertices $i$ and $j$ that belong to the same community, this difference is expressed as
\begin{equation}
\frac{1}{2} \sum_{i,j} ^{N}\left( A_{ij} - \frac{k_i k_j}{2m} \right) \delta(c_i, c_j),
\end{equation}
where $N$ is the total number of vertices and $A_{ij}$ is the element of the adjacency matrix ($A_{ij}=A_{ij}=1)$. The Kronecker delta $\delta(c_{i}, c_{j}) = 1$, when $i$ and $j$ are in the same community; else 0, conventionally one calculates not the number of such edges, but the fraction, which is given by the same expression divided by the number $m$ of edges:
\begin{equation}
Q = \frac{1}{2m} \sum_{i,j} \left( A_{ij} - \frac{k_i k_j}{2m} \right) \delta(c_i, c_j).
\label{modularity}
\end{equation}
This quantity $Q$ is called modularity \cite{newman2004finding,newman2003mixing} and is a measure of the extent to which the vertices are connected in the network. It is always $Q < 1$. It is $Q > 0$ if there are more edges between vertices of the same type than we would expect by chance, and $Q < 0$ if there are fewer. The modularity of a weighted network, as introduced in~\cite{newman2004analysis}, extends the classical definition to account for the intensity of connections between nodes. It is expressed as
\begin{equation}
Q = \frac{1}{2M} \sum_{i,j} \left( w_{ij} - \frac{s_i s_j}{2M} \right)\delta(c_i, c_j),
\end{equation}
where $w_{ij}$ denotes the weight of the edge between vertices $i$ and $j$ and 
\(
M = \frac{1}{2}\sum_{i=1}^{N}\sum_{j=1}^{N} w_{ij}
\)
represents the total weight of all edges in the network. Here, the usual notion of degree is replaced by the \emph{strength} of a vertex. For a node $i$, this strength is given by
\(
s_i = \sum_{l=1}^{N} w_{li},
\)
capturing the cumulative weight of all connections incident to the node.
Over the years, researchers have proposed a rich collection of community detection algorithms, each built on a distinct intuition about how \textit{communities} emerge in complex systems. These approaches range from hierarchical clustering and modularity-driven optimization to label-propagation schemes and spectral techniques. Every class of methods offers its own strengths and insights, helping reveal different layers of structure hidden within real-world networks. In the following subsections, we present the mathematical foundations of a selection of well-known algorithms representing these diverse perspectives.
\subsection{Algorithms for Detecting \textit{Community Structure} in Networks}
\justifying 
The Girvan--Newman algorithm~\cite{newman2004finding}, introduced in the early 2000s, is one of the earliest systematic approaches for uncovering \textit{community structure} in networks. Unlike agglomerative methods that build \textit{communities} upward from individual vertices, this algorithm follows a divisive philosophy: it begins with the entire network. It gradually dismantles it by removing edges that appear to act as critical links between groups. At the heart of the method lies the notion of edge betweenness centrality \cite{newman2018networks}, which counts the number of shortest paths in the network that pass through a given edge. Edges with high betweenness often play the role of bridges connecting otherwise well-knit regions, and removing them tends to expose the underlying modular structure. The algorithm proceeds through the following steps: (i) Start by calculating the betweenness centrality for each edge in the network.
(ii) Then, remove the edge that has the highest betweenness value.
(iii) Once the network is updated, compute the betweenness centrality again.
(iv) Keep repeating the removal and recalculation steps until no edges remain.

By repeatedly removing the edges most responsible for inter-community connectivity, the algorithm effectively separates the network into progressively smaller, internally cohesive groups. At each stage, the resulting partition can be evaluated using modularity $Q$, allowing one to identify the division that best captures the network's inherent structure. Although the Girvan--Newman algorithm offers clear interpretability and often reveals meaningful hierarchical organization, its reliance on recomputing betweenness at every iteration makes it computationally demanding for large-scale networks. Even so, it remains a foundational technique in community detection and continues to serve as a conceptual cornerstone for many later developments in the field.

Further in the same year (2004), Newman~\cite{Newman2004} introduced a faster variant of the classical community detection approach. This method incrementally builds community partitions and evaluates their quality at every step using the modularity measure \(Q\). The modularity in this formulation is given by
\begin{equation}
Q = \sum_{i} \big[\, L_{ii} - a_i^{\,2} \,\big],
\label{Eq.18}
\end{equation}
where \(L_{ij}\) denotes the fraction of all network edges that run between \textit{communities} \(i\) and \(j\) and 
\(
a_i = \sum_{j} L_{ij}
\)
represents the total fraction of edges connected to community \(i\). Both \(L_{ij}\) and \(a_i\) naturally incorporate edge weights whenever the network is weighted. A detailed discussion of this modularity expression can be found in~\cite{Newman2004}.

The algorithm begins with the most fine-grained possible partition: each of the \(N\) vertices is treated as its own community. The method then considers all pairs of \textit{communities} and evaluates how the modularity would change if they were merged. For a pair of \textit{communities} \(i\) and \(j\), the corresponding change in modularity is computed as
\begin{equation}
\Delta Q = L_{ij} + L_{ji} - 2 a_i a_j.
\end{equation}
At each iteration, the pair with the highest positive \(\Delta Q\) is merged, since this creates the greatest improvement in modularity. The process continues in this greedy fashion until no further merge leads to a meaningful increase in \(Q\). Finally, the algorithm produces the optimal \textit{community structure} of the network that maximizes the modularity metric. The final hierarchy produced by these successive merges often reveals a clear and interpretable \textit{community structure} within the network. Although built on a simple greedy principle, this algorithm captures the essence of modularity-driven community detection with remarkable speed and clarity. Its ability to uncover meaningful groups in networks that were previously too large to analyse makes it a reliable baseline for practical applications. At the same time, its reliance on local merge decisions highlights a natural trade-off: impressive efficiency, but occasionally at the cost of the finer structural details that more global methods may capture.

In the year (2006), Newman \cite{newman2006modularity} extended and proposed one of the most influential community detection approaches is based on maximizing the modularity function $Q$, which measures the quality of a network partition relative to a randomized null model~\cite{newman2006finding}. For an undirected, unweighted network, whose adjacency matrix is $A$ and degree sequence is $\{k_i\}$, the modularity of a partition $\{g_i\}$ is
\begin{equation}
    Q = \frac{1}{2m} 
        \sum_{i,j} \left(A_{ij} - \frac{k_i k_j}{2m}\right) 
        \delta(g_i, g_j),
\end{equation}
where $m = \frac{1}{2}\sum_{ij} A_{ij}$ is the total connections in network and $\delta(g_i, g_j)=1$ if vertices $i$ and $j$ belong to the same community,  and $0$ otherwise.

\noindent
Introducing the modularity matrix
\begin{equation}
    B_{ij} = A_{ij} - \frac{k_i k_j}{2m},
\end{equation}
The modularity can be expressed as 
\begin{equation}
    Q = \frac{1}{4m}\,\mathbf{s}^{T} B\,\mathbf{s},
\end{equation}
where $\mathbf{s}$ is an $n$-dimensional vector with components $s_i=\pm1$ indicating the community assignment in a bi-partition. Maximizing $Q$ is NP-hard, but an approximate and effective solution is obtained  by relaxing $\mathbf{s}$ to take continuous values and aligning it with the  leading eigenvector $\mathbf{u}_1$ of $B$:
\begin{equation}
    s_i = 
    \begin{cases}
        +1, & u_{1i} \geq 0,\\
        -1, & u_{1i} < 0.
    \end{cases}
\end{equation}
This spectral method recursively bisects \textit{communities} until no further subdivision increases $Q$. The significance of this method lies in its elegant connection between the network structure and spectral properties: the eigenvalues of $B$ quantify the strength of modular organization, while eigenvectors guide optimal partitions. This approach is computationally efficient, scalable to large networks and has become a cornerstone in network science, enabling applications in social, biological and technological systems.

Blondel et al.~\cite{blondel2008fast} introduced a widely used modularity-based method in (2008), commonly known as the Louvain algorithm. The appeal of this approach lies in its simplicity and speed: it follows a greedy strategy to increase the modularity \(Q\) (defined in Eq.~\ref{modularity}) and in practice, runs in roughly \(O(n \log n)\), making it suitable for very large networks. The algorithm proceeds in two alternating phases and these phases are repeated until the \textit{community structure} no longer changes. In Phase one, at the outset, every vertex forms its own community. The algorithm then visits each node \(i\) and considers moving it into the community of one of its neighbors. For each potential move, the change in modularity \(\Delta Q\) is computed and the move is adopted whenever it leads to an improvement. This simple reassignment step is repeated until no further gain in modularity is possible. The modularity gain achieved by moving a node \(i\) into a community \(U\) is
\begin{equation}
\Delta Q 
=
\left[
\frac{\sum_{\text{in}} + k_{i,\text{in}}}{2m}
-
\left( \frac{\sum_{\text{tot}} + k_i}{2m} \right)^{2}
\right]
-
\left[
\frac{\sum_{\text{in}}}{2m}
-
\left( \frac{\sum_{\text{tot}}}{2m} \right)^{2}
-
\left( \frac{k_i}{2m} \right)^{2}
\right],
\end{equation}
where \(\sum_{\text{in}}\) is the total weight of internal links in \(U\), \(\sum_{\text{tot}}\) is the total weight of edges touching nodes in \(U\), \(k_i\) is the weighted degree of \(i\), \(k_{i,\text{in}}\) is the sum of weights from \(i\) to nodes in \(U\) and \(m\) is the total edge weight in the network. This phase yields a partition that represents a local maximum of modularity.

\noindent 
Once the first phase stabilizes, in the second phase, the discovered communities are collapsed into super-nodes to form a new weighted network. Two communities \(U\) and \(V\) are connected in this reduced network if any node in \(U\) has an edge to a node in \(V\). The weight between them is simply the sum of all such edges:
\[
W_{UV} = \sum_{i \in U}\sum_{j \in V} A_{ij}.
\]
Similarly, each community carries a self-loop with weight
\[
W_{UU} = \sum_{i \in U}\sum_{j \in U} A_{ij}.
\]
The algorithm then returns to Phase~I on this reduced network. By repeating these two phases, the Louvain method naturally uncovers \textit{communities} at multiple scales and produces a hierarchical representation of the network.

\noindent
In essence, the Louvain method turns modularity optimization into an intuitive, multi-level exploration, allowing \textit{communities} to emerge from fine detail to broader structure with remarkable ease. Its ability to scale to millions of nodes without sacrificing quality has made it a trusted tool in the study of large, real-world networks. At the same time, its hierarchical nature invites deeper insight, reminding us that \textit{community structure} is rarely flat and that meaningful patterns often reveal themselves layer by layer.

The modularity-based methods give us a global lens on \textit{community structure}, but networks can also be understood through the small, repeated interactions that happen between nodes. This is where label propagation algorithms come in, the first ground method, the asynchronous label propagation algorithm (ALPA). Raghavan et al.\ (2007) \cite{raghavan2007near} introduced a simple yet strikingly effective community detection method based purely on local consensus formation.  The algorithm relies on the idea that a node tends to join the group most common among its immediate neighbours, allowing \textit{communities} to emerge naturally from repeated label interactions.  No objective function is optimized and no prior information about the number or size of \textit{communities} is required, making the process entirely structure-driven.

Let \(G=(V,E)\) be an undirected network with adjacency list \(\Gamma(i)\) for each node \(i\). Every node initially receives a unique label. Assign each node a unique identifier:
\[
C_i(0) = i, \qquad \forall i \in V.
\]
At iteration \(t\), choose a random permutation \(X\) of all nodes. Nodes are updated sequentially following this order. For a node \(i\) at iteration \(t\), let its neighbours be
\[
\Gamma(i) = \{j_1,\ldots,j_{k_i}\}.
\]
Define \(d^l_i(t)\) as the number of neighbours currently carrying label \(l\). Node \(i\) adopts the most frequent label in its neighborhood:
\begin{equation}
C_i(t) = \arg\max_{l}\, d^l_i(t),
\end{equation}
with ties broken uniformly at random. Already-updated neighbours use iteration \(t\) labels, others use iteration \(t\!-\!1\). The process continues until every node carries a label that is at least as common among its neighbours as any other label:
\[
C_i = l \quad \Longrightarrow \quad d^l_i \ge d^{l'}_i \ \ \forall l'.
\]
Nodes with identical labels form \textit{communities}. The final partition is
\[
\mathcal{C} = \{\, C_1, C_2, \ldots, C_K \}, 
\qquad
C_k = \{\, i : C_i = k \,\}.
\]
The algorithm operates in near-linear time \(O(m)\) per iteration, making it suitable for very large networks.  It's purely a local decision rule that reveals \textit{communities} as emergent structures, rather than as solutions to a global optimization problem.  The asynchronous updates overcome oscillations seen in synchronous schemes, particularly
in bipartite or star-like regions.  Although different runs may yield different valid partitions, these solutions often agree substantially and reflect meaningful alternatives in networks with overlapping structure.  Overall, the method provides a fast, intuitive and remarkably scalable route to discovering \textit{communities} in complex real-world graphs.

Later, Lou et al.~\cite{lou2013detecting} improved the classical Label Propagation Algorithm (LPA) by replacing the naïve ``most-frequent neighbour label'' rule with a weighted coherent-neighborhood propinquity (weighted-CNP) measure that quantifies the probability that a vertex and a neighbour belong to the same community. Let $G=(V, E)$ be a simple, possibly weighted network with edge weights $w_{uv}>0$ for $(u,v)\in E$.  The classical Label Propagation Algorithm (LPA) assigns an initial and unique label to every vertex and then repeatedly updates each label by adopting the most frequent label among its neighbours.  Although efficient, this approach treats all neighbours uniformly and hence may overlook subtle structural patterns in the local topology. To address this limitation, the method introduces a quantitative measure called the weighted coherent neighbourhood propinquity (wCNP).  The purpose of this measure is to capture how strongly two vertices ``tend'' to participate in the same community by examining the coherence of their neighbourhoods and the strengths of the edges connecting them.  For adjacent vertices $u$ and $v$, the wCNP value is defined as
\begin{equation}
    \mathrm{wCNP}(u,v)
      = 
      \frac{
          \displaystyle
          \sum_{x \in N(u)\cap N(v)} 
          \phi\!\left(w_{ux},\, w_{vx}\right)
      }{
          \displaystyle
          \sum_{y \in N(u) \cup N(v)}
          \psi\!\left(w_{uy},\, w_{vy}\right)
      },
\end{equation}
where $N(u)$ denotes the set of neighbours of $u$ and $\phi$ and $\psi$ are smooth, monotonic functions that quantify local coherence (for example, $\phi(a,b)=ab$ and $\psi(a,b)=a+b$).  The numerator captures the strength of the shared neighbourhood between $u$ and $v$, while the denominator ensures proper normalization.

\medskip
To discount vertices whose neighbourhood labels fluctuate widely, the method incorporates a simple entropy-based adjustment. For a vertex $v$, let $p_\ell^{(v)}$ be the proportion of neighbours of $v$ carrying label $\ell$ at iteration $t-1$.  The local label entropy is then
\begin{equation}
    H(v) = -\sum_{\ell} p_\ell^{(v)}\log p_\ell^{(v)}.
\end{equation}
Vertices with ambiguous neighbourhoods (high $H(v)$) receive lower influence during propagation.  Thus, the entropy-adjusted propinquity becomes
\begin{equation}
    \widetilde{\mathrm{wCNP}}(u,v)
        = \mathrm{wCNP}(u,v)\,
          \bigl(1-H(u)\bigr)\bigl(1-H(v)\bigr).
\end{equation}

\medskip
The label update rule is then modified to incorporate this refined local information.  For each vertex $v$ at iteration $t$, one sets
\begin{equation}
    \ell_v(t)
    =
    \arg\max_{\ell}
    \sum_{\substack{u \in N(v)\\[2pt]
                    \ell_u(t-1)=\ell}}
         \widetilde{\mathrm{wCNP}}(v,u).
\end{equation}
In contrast with classical LPA, where each neighbour contributes equally, the updated rule assigns a larger contribution to neighbours that are coherently aligned with $v$ within the local topology. Viewed from a community-detection perspective, this formulation guides the propagation process toward labels that are not merely abundant in 
the immediate neighbourhood, but that are also structurally coherent and topologically meaningful.  Empirical results reported in the literature show that this weighted and entropy-modulated variant produces more stable partitions and higher quality \textit{community structure} while preserving the low computational 
overhead characteristic of LPA. This refinement of label propagation enriches the local structure of the decision-making process by allowing each neighbour to influence a vertex in proportion to their coherent neighbourhood similarity.  As a result, community detection becomes noticeably more stable and less dependent on random update orderings. The method preserves the simplicity of classical LPA while achieving consistently higher-quality partitions in real-world networks.

Further, Zhang et al.~\cite{zhang2017label} have refined the classical label--propagation framework by integrating two complementary ideas: the intrinsic importance of each node and the influence carried by the labels present in its neighbourhood. For a given undirected graph \(G=(V, E)\) with \(n=|V|\) nodes, this method initially assigned a unique label to each node, so that
\begin{equation}
c_i(0)=i, \qquad i\in V,
\end{equation}
which ensures that no prior assumptions about the \textit{community structure} are imposed.  This simple starting point lets the evolution of labels be driven
entirely by local interactions and influence. Although every node starts with its own label, not every node plays an equally central role in the network.  This introduces a numerical measure \(\operatorname{Inf}(i)\) describing the prior importance of node \(i\), obtained from behavioural attributes or other external information.  However, a node's position in the graph also contributes to its significance.  To capture this, it refines the importance score by incorporating the importance of its neighbours:
\begin{equation}
\operatorname{NI}(i) = \operatorname{Inf}(i) + \alpha \sum_{j\in N(i)} \frac{\operatorname{Inf}(j)}{d(j)}
\end{equation}
where \(0\le \alpha\le 1\) controls how strongly neighbouring nodes contribute to \(i\)'s overall influence.  Nodes are then ordered in decreasing \(\operatorname{NI}(\cdot)\), which anchors the subsequent propagation process and removes the randomness present in classical label--propagation.

With the node order fixed, next update labels iteratively. At iteration \(t\), nodes are processed in the sequence
\begin{equation}
x_1,\, x_2,\, \ldots,\, x_n
\quad\text{satisfying}\quad
\operatorname{NI}(x_1)\ge \cdots \ge \operatorname{NI}(x_n).
\end{equation}
This choice ensures that nodes deemed more influential propagate their labels earlier, allowing them to shape the emerging \textit{community structure}.  For each node \(i\), inspect the labels present in its neighborhood.  Let \(f(l)\) denote the number of neighbours of \(i\) carrying label
\(l\).  The labels achieving the maximum support are
\begin{equation}
L_{\max}(i)=
\bigl\{l \,\mid\, f(l)=\max_{l'} f(l') \bigr\}.
\end{equation}
If this set contains only one label, the update is immediate; otherwise, resolve the tie by examining label influence. When several labels are equally represented around node \(i\), their respective influence is not identical.  To quantify this, define for each candidate label \(l\) the quantity
\begin{equation}
\operatorname{LI}(i,l)
=
\sum_{j\in N_l(i)} 
\frac{\operatorname{NI}(j)}{d(j)}
\end{equation}
where \(N_l(i)\subseteq N(i)\) is the set of neighbours of \(i\) that currently carry label \(l\). This measure reflects the idea that a label should exert more pressure on \(i\) if it is held by neighbours that are themselves influential. The updated label of \(i\) is thus chosen as
\begin{equation}
c_i(t)
=
\arg\max_{l\in L_{\max}(i)} \operatorname{LI}(i,l)
\end{equation}
This rule eliminates the randomness of traditional LPA and produces a stable, influence-driven evolution of labels. The iterative updates continue until either a maximum number of iterations is reached or the labels stabilize, that is,
\begin{equation}
c_i(t)=c_i(t-1)
\qquad\text{for all } i\in V.
\end{equation}
When the process converges, all nodes sharing the same final label are grouped into a single community:
\begin{equation}
C_l = \{\, i \in V \; :\;  c_i = l \,\}.
\end{equation}
The collection of these sets forms the resulting \textit{community structure}. By combining node importance, deterministic update order and label--influence selection, the proposed method yields \textit{communities} that are both stable and coherent with the intrinsic and structural properties of the network. This approach brings stability and clarity to label propagation by letting the most influential nodes guide how \textit{communities} form, rather than leaving the process to chance. By blending prior node importance with the strength of label influence, the method uncovers groups that genuinely reflect how information and authority flow through the network. As a result, the detected \textit{communities} are more consistent, interpretable and aligned with the actual structure of large, complex systems.

In this line, Berahmand and Bouyer (2018) \cite{berahmand2018lp} introduced LP (Local Path)-LPA to stabilise classical Label Propagation (LPA) by replacing random decisions with a principled, semi-local similarity strategy.  The algorithm assigns update priorities based on the label influence of nodes and selects propagated labels using link strength rather than majority voting.  This removes instability, eliminates tie–break randomness and prevents monster-community formation. Let \(G = (V, E)\) be an undirected network with adjacency matrix \(A\). Each node initially receives a unique label.

\noindent
LP--LPA uses the Local Path (LP) similarity, which incorporates paths of length 1--3 and balances accuracy with low computational cost:
\begin{equation}
S(u,v)=\beta A_{uv} + \beta^{2} (A^{2})_{uv} + \beta^{3} (A^{3})_{uv},
\qquad \beta = \frac{1}{d_{\mathrm{mean}}}.
\end{equation}
This semi-local similarity reflects the connectivity strength within two or three hops, capturing the typical diameter of \textit{communities} in real networks.

\noindent
For each node \(u\), label influence is defined as the aggregated similarity with its immediate neighbours:
\[
k(u) = \sum_{v\in \Gamma_1(u)} S(u,v).
\]
Nodes with high influence are treated as \emph{cores}, while hubs and bridge nodes naturally receive lower influence and therefore update later.  Nodes are sorted in descending order of \(k(u)\), forming the deterministic update order. When node \(u\) updates its label, it selects the label of the neighbour \(v\) that has the strongest link with it:
\begin{equation}
\mathrm{LS}(u) = 
\arg\max_{v\in\Gamma_1(u)} S(u,v).
\end{equation}
If multiple neighbours share the same maximum link strength, their strengths are summed label-wise and the label with the highest cumulative strength is chosen.  If ties persist, the node retains its current label, ensuring deterministic behaviour.

\noindent
Let \(t\) denote the iteration index.  LP--LPA repeatedly scans nodes in the influence-based order:
\begin{equation}
\text{for } u\in X:\quad
\ell_u^{(t+1)} = \text{label determined by link strength rule}.
\end{equation}
Propagation stops when no node changes its label or when a small maximum number of iterations is reached (typically 3--5 due to rapid convergence).
\noindent
Nodes sharing identical labels after convergence form the final \textit{communities}:
\begin{equation}
C = \{C_1,\dots,C_K\}, \qquad
C_k = \{u\in V : \ell_u = k\}.
\end{equation}
LP-LPA transforms LPA from a random and unstable procedure into a deterministic, semi-local, influence-guided propagation method.  Its use of similarity-based influence correctly distinguishes core, hub and bridge nodes, preventing label leakage across \textit{communities}.  Practically, the method offers near-linear running time while maintaining high accuracy on large networks.  Mathematically, it embeds structural similarity into both update order and label selection, tightly coupling propagation behaviour with network topology.  Empirically, LP--LPA consistently avoids monster \textit{communities} and achieves high stability across runs.

One other popular framework, known as Spectral methods, offers a remarkably intuitive way to uncover \textit{communities}: they translate the structure of a network into the geometry of its eigenvectors, allowing clusters to appear as natural patterns in a low-dimensional space. Instead of working directly on raw connections, these approaches let the network “speak” through its spectra, revealing subtle relationships that are often invisible at the surface. Because of this ability to capture both global structure and fine-scale organization, spectral clustering has become one of the most insightful tools in modern network analysis. The following methods illustrate how different spectral formulations enrich this idea of network community detection approaches. 

Xie et al.\ (2009) \cite{xie2009detection} proposed a thoughtful refinement of spectral community detection by strengthening the notion of node similarity before applying the spectral machinery.  The method begins by enriching the classical shared–nearest–neighbour (SNN) similarity, projects the network into a spectral space and then lets Fuzzy \(c\)-Means (FCM) guide the final grouping.  This combination yields a practical and flexible framework, especially suited for networks whose \textit{communities} are present but not sharply separated. Let \(G=(V,E)\) be an undirected graph with adjacency matrix \(\mathbf A\), neighbour set \(\Gamma(i)\) and degree \(d_i = |\Gamma(i)|\).
\noindent
The basic SNN similarity between nodes \(i\) and \(j\) counts how many neighbours they 
share:
\begin{equation}
S_{ij} = |\Gamma(i)\cap \Gamma(j)|.
\end{equation}
While informative, this measure weighs all nodes equally.  To capture how meaningful the shared neighbours are to each endpoint, the authors introduce an asymmetric refinement:
\begin{equation}
S'_{ij} =
\begin{cases}
\dfrac{S_{ij} A_{ij}}{d_i}, & i\neq j,\\[4pt]
0, & i=j.
\end{cases}
\end{equation}
A row-normalised version,
\begin{equation}
S'' = K^{-1} S', \qquad
K_{ii} = \sum_j S'_{ij},
\end{equation}
yields a stochastic matrix whose dominant eigenvalue is \(1\).  
The remaining eigenvalues encode graded structural information that can reveal 
community tendencies.
\noindent
Compute the eigenvalues and eigenvectors of \(S''\):
\begin{equation}
1 = \lambda_1 > \lambda_2 > \cdots > \lambda_k.
\end{equation}
The eigenvector associated with \(\lambda_2\) captures the strongest nontrivial mode of variation in the network and serves as the first one-dimensional embedding of all nodes.
\noindent
Apply Fuzzy \(c\)-Means (FCM) to the selected spectral coordinates.  The resulting soft assignments reflect how strongly each node aligns with each potential community. For every obtained clustering, compute the modularity \(Q\), defined in Eq. ~\ref{modularity}, which serves as the global indicator of community quality.

To capture subtler structures, include the next leading eigenvector (\(\lambda_3\)),  Repeat the FCM clustering and recompute \(Q\).  Continue growing the feature set step by step:
\[
\{\mathbf u_2\},\;
\{\mathbf u_2,\mathbf u_3\},\;
\{\mathbf u_2,\mathbf u_3,\mathbf u_4\}, \dots
\]
This iterative enrichment allows the method to explore different degrees of 
spectral resolution. The best community division is the one that obtains the largest modularity value:
\begin{equation}
\mathcal C^\ast = \arg\max_{\mathcal C} Q(\mathcal C).
\end{equation}
This improved spectral framework strengthens the input to spectral analysis by incorporating a degree-aware, shared-neighbour similarity that reflects genuine local cohesion.  The fusion of spectral embedding with FCM provides a smooth way to navigate ambiguous or overlapping boundaries between \textit{communities}. Modularity then offers a principled criterion to select the most coherent cluster structure.  Overall, this method elegantly connects similarity enhancement, spectral geometry and fuzzy clustering, producing reliable results even when community boundaries are not sharply defined.

Then, in the year (2020), Xu proposed \cite{xu2020spectral} a community–detection approach that blends communicability measures with a spectral relaxation framework. The proposed spectral framework detects \textit{communities} by comparing how strongly nodes communicate in the real network versus how strongly they would be expected to communicate in a suitable null model.  The method replaces the traditional edge-based modularity with a communicability-based criterion, allowing longer walks and multi-step influence to shape the detection of \textit{community structure}.  The resulting modularity matrix admits a natural spectral decomposition, from which a community partition is obtained through a vector-partitioning procedure. Let \(A\) be the adjacency matrix of an undirected network \(G=(V,E)\),  and let \(e^{A}\) denote its communicability matrix with entries
\begin{equation}
(e^{A})_{pq} = \sum_{k=0}^{\infty} \frac{(A^{k})_{pq}}{k!}.
\end{equation}
The total communicability is
\begin{equation}
TC(G) = \sum_{p,q} (e^{A})_{pq}.
\end{equation}
The null model used for comparison is based on expected degree contributions,
\[
P_{pq} = \frac{d_p d_q}{2m},
\]
and its communicability is
\[
(e^{P})_{pq} = \sum_{k=0}^{\infty} \frac{(P^{k})_{pq}}{k!}.
\]
Define the modularity matrix
\begin{equation}
B_{pq} = (e^{A})_{pq} - (e^{P})_{pq},
\end{equation}
so that for a partition \(\{c_p\}\),
\begin{equation}
Q_2 = \frac{1}{TC(G)} \sum_{p,q} B_{pq}\,\delta(c_p,c_q).
\end{equation}
Since \(B\) is symmetric, write its eigen-decomposition
\begin{equation}
B_{pq} = \sum_{l=1}^{n} \lambda_l U_{pl}U_{ql}, \qquad
\lambda_1 \ge \lambda_2 \ge \cdots \ge \lambda_n.
\end{equation}
Keep only the \(i\) largest positive eigenvalues. Define for each vertex \(p\) an \(i\)-dimensional vector
\begin{equation}
\mathbf{r}_p = \bigl(\sqrt{\lambda_1}U_{p1},\,\sqrt{\lambda_2}U_{p2},\,\dots,\sqrt{\lambda_i}U_{pi}\bigr).
\end{equation}
For a \(k\)-community division, let each community \(s\) have a group vector
\begin{equation}
\mathbf{R}_s = \sum_{p\in s} \mathbf{r}_p.
\end{equation}
The approximate modularity becomes
\begin{equation}
Q_2 \approx \frac{1}{TC(G)} \sum_{s=1}^k \|\mathbf{R}_s\|^2.
\end{equation}
Move a vertex \(p\) from community \(s\) to \(t\) if it increases the modularity:
\begin{equation}
\Delta Q = \frac{2}{TC(G)}
\left(\mathbf{R}_t^\top \mathbf{r}_p - \mathbf{R}_s^\top \mathbf{r}_p\right) > 0.
\end{equation}
Update all group vectors and repeat the reassignment until convergence.

\noindent
Because the result depends on initial group vectors, the algorithm is run multiple times with random initializations.  The partition achieving the largest \(Q_2\) is selected as the final \textit{community structure}. This approach extends modularity to incorporate all possible walks, allowing subtle structural affinities to influence community formation.  Its spectral embedding provides a clean geometric interpretation of communicability patterns inside the network.  The vector-partitioning stage is computationally light, giving the method good practical scalability.  Because it naturally amplifies dense communication within groups, the algorithm excels at identifying tightly knit modules in both real and synthetic networks.  
Overall, the method provides a refined and globally informed perspective on community detection that goes beyond traditional edge-based modularity.

Taken together, the modularity-based methods, label-propagation approaches and spectral techniques each shed light on \textit{communities} from a different angle, offering complementary strengths. Modularity maximisation captures global structure well, but often struggles with resolution limits in finely partitioned networks. Label propagation brings impressive speed and scalability, yet its simplicity can leave it sensitive to instability without careful refinement.
Spectral decomposition provides a rich geometric view of the network, though it can be computationally heavier on very large graphs. Understanding these trade-offs helps us appreciate why no single method is universally superior—and why blending ideas across these families continues to inspire more robust and insightful community-detection algorithms.

\subsection{Exploitation of \textit{Community Structure} Information for Network Analysis}

\textit{community structure} are fundamental to understanding the organization and functionality of complex networks. Several influential studies have leveraged community information to deepen insights into network dynamics and node behavior. For instance, \cite{guimera2005functional} introduced the concepts of within-community degree and participation coefficient to classify nodes into distinct structural roles such as hubs, connectors and peripherals. This pioneering framework provides a foundation for analyzing functional roles based on the interplay between local connectivity and global community organization. Building upon the temporal dimension of network structure, \cite{delvenne2010stability} proposed a Markov stability framework that defines a time-scale–dependent stability measure for clustering. By modeling diffusion processes over the network, their approach revealed multi-resolution \textit{community structure} and demonstrated the robustness and interpretability of community partitions across varying dynamical scales.

In another direction, \cite{palla2005uncovering} developed the Clique Percolation Method (CPM) to detect overlapping \textit{communities} and applied it to biological and social systems. Their method uncovered hierarchical patterns of community overlap and introduced novel measures to characterize structural redundancy and inter-community relationships. Further extending the application of community information, \cite{zheleva2008using} incorporated detected social groups—such as friendship and family circles—into a probabilistic model for link prediction. Their findings showed that community-based features substantially improve prediction accuracy compared to models that rely solely on direct pairwise relationships. Collectively, these studies illustrate the diverse ways in which \textit{community structure} information can be exploited for network analysis—from identifying functional roles and uncovering multi-scale modular patterns to improving predictive modeling and understanding overlapping structures.

\subsection{Applications of \textit{Community Structure} in Real-World}

The identification of \textit{community structure} is fundamental to uncovering the functional organization of complex systems, providing critical insights into their generative mechanisms, dynamical behavior and resilience properties. In real-world complex systems, community detection elucidates patterns of interaction and influence propagation, enabling quantitative modeling of opinion dynamics, information spread and the formation of social capital. Thus, Community detection plays a pivotal role in understanding the mesoscopic organization of complex systems and has found wide-ranging applications across diverse real-world domains. Applications of network \textit{communities} in modeling real-world systems have explored biological, social, financial and communication aspects. Here, we review some of the relevant community applications in the above field.

In biological and biomedical networks, one of the early works that identified densely connected clusters in PPI networks interprets them as protein complexes or dynamic functional modules. \textit{communities} (modules) map to functional units in the cell \cite{spirin2003protein}. In the same year, Tornow \cite{tornow2003functional} integrated PPI networks with gene expression (co‐expression) networks, using module/community detection to find functional modules that are coherent both topologically and in expression space. Later in the year 2006, Chen \cite{chen2006detecting} developed a partitioning algorithm (community detection) on a weighted yeast PPI network (using functional genomics weights) and showed that the identified modules correspond to phenotypic similarity, known complexes, etc. 

Further, Lewis, in the year 2010 \cite{lewis2010function}, explored how \textit{community structure} in a yeast PPI network behaves at different resolution scales and how \textit{communities} correspond to functionally homogeneous modules. In a similar year \cite{zhang2010determining}, used a community detection method (maximizing modularity density) to identify functional modules in PPI networks; they used modularity‐based detection and showed that the modules have biological significance. Then, Sah in the year 2014 \cite{sah2014exploring} applied community detection on empirical biological networks, benchmarked algorithms via modular random graphs and explored how \textit{community structure} emerges in biological systems.
Further research in the year 2017 \cite{pesantez2017efficient} focuses on community detection in bipartite biological networks (two types of nodes, e.g., genes \& diseases, hosts \& pathogens), extending classic methods to this setting for biologically‐relevant clustering. A Comparative paper  \cite{rahiminejad2019topological} that applies different community detection algorithms to yeast \& human PPI networks, assesses which methods yield biologically meaningful modules (GO/KEGG enrichment). A recent review \cite{manipur2021community} of community detection methods in PPI networks, summarizing how module detection is applied for disease‐mechanism identification, biomarker detection, etc. Thus community's application in the field of biology gives meaningful patterns and insights about the systematic behavior of the system.

On the other hand, in social networks, identifying \textit{communities} enables the analysis of friendship groups, professional circles and interest-based clusters, which is valuable for targeted marketing, recommendation systems and information diffusion modeling. The application of community detection in social systems has provided deep insights into how human interactions and relationships organize into cohesive structures. Foundational studies by Girvan and Newman \cite{girvan2002community,newman2004finding} first revealed modular patterns in collaboration and friendship networks, establishing modularity as a key measure of social cohesion. Subsequent work by Raghavan \textit{et al.} (2007) \cite{raghavan2007near} introduced efficient label propagation methods to detect \textit{communities} in large-scale online social networks, uncovering natural groupings such as friendship circles and professional clusters. With the rise of dynamic and multilayer data, Harenberg \textit{et al.} (2014) \cite{harenberg2014community} and Rossetti \textit{et al.} (2017) \cite{rossetti2017tiles} advanced techniques to track temporal evolution and multiplex \textit{community structure} across communication and digital platforms. More recent studies, such as Shi \textit{et al.} (2017) \cite{shi2017community} and Fortunato and Hric (2016) \cite{fortunato2010community}, integrated behavioral and influence-based factors, showing that community cores often correspond to socially influential individuals or ideologically aligned groups. Collectively, these works demonstrate that community detection has become a vital tool for mapping social cohesion, influence diffusion and collective behavior in complex, interconnected societies.

The application of community detection techniques to financial systems has grown steadily in both breadth and sophistication. Early work by L. Bargigli and M. Gallegati (2013) \cite{bargigli2013finding} analyzed a bipartite bank-firm credit network, demonstrating that banks and firms cluster into tightly‐connected modules, thereby providing a meso-level structure relevant for contagion risk. This was followed by the study by J. A. Chan‑Lau (2018) \cite{chan2018systemic} in Quantitative Finance and Economics introduced “systemic \textit{communities}” identified via edge-betweenness and the map-equation (Infomap) in a global financial variance‐decomposition network, showing that community detection complements centrality measures for assessing systemic importance. Moving into the asset‐allocation sphere, L. Zhao \textit{et al.} (2021) \cite{zhao2021community} employed correlation‐based stock networks to detect \textit{communities} of securities and used those clusters for portfolio construction, finding improved diversification benefits.

In 2021, Yajing Huang and Feng Chen \cite{huang2021community} applied community detection to U.S. bank correlation networks (derived from investment portfolios), showing that smaller \textit{communities} often bear much higher systemic risk compared to large, more stable modules. The portfolio‐allocation framework by S. Ferretti (2023) \cite{ferretti2023modeling} built on this by structuring asset allocation around modular \textit{communities} rather than traditional covariance alone, emphasizing diversification across network modules. Most recently, a study \cite{yang2023communities} examined co‐occurrence news‐based networks of financial firms, used Louvain community detection to identify firm clusters with dense inter-connections and linked these to systemic risk indicators. Collectively, this selection of studies traces the evolution of community detection in finance—from structural credit networks, through systemic‐risk identification and asset‐allocation, to text-driven network clustering—and underscores how mesoscopic network modules have become a versatile tool for understanding, monitoring and managing complexity in financial systems. Additionally, in the financial domain, \textit{community structure} have also been effectively applied to design portfolio investment strategies and to track market dynamics by examining the quality and group patterns of these \textit{communities} \cite{zhao2021community, pawanesh2025exploring, nie2018constructing, pawanesh2025exploiting}.

In technological and infrastructural systems, the identification of densely connected substructures improves network design, facilitates fault-tolerant architectures and enhances the efficiency of information routing and resource allocation \cite{clauset2004finding,fortunato2016community}. In economic and financial networks, community detection reveals market segmentation, inter-dependencies between assets or institutions and hidden channels of systemic risk transmission, thereby informing macro-prudential regulation and portfolio diversification strategies \cite{fenn2009dynamic,barucca2016disentangling,bargigli2013finding}. In transportation and mobility networks, modular partitions expose latent geographic or operational regions, which can optimize traffic management, urban planning and robustness against cascading failures \cite{guimera2005worldwide,derrible2012network}. These applications demonstrate that \textit{community structure} is not merely a descriptive property but a quantitatively significant feature, providing a unifying framework for analyzing, interpreting and engineering the structural and functional dynamics of real-world complex systems.

\subsection{Significance of Network \textit{Community Structure}}

\textit{community structure} provides a mesoscopic perspective on complex networks, bridging the gap between local node-level interactions and global macroscopic organization. Mathematically, \textit{communities} enable coarse-graining networks into simplified representations that preserve key topological characteristics while facilitating the computation of higher-order measures, such as entropy, modularity and stability. In real-world systems, these \textit{communities} often correspond to functionally meaningful groupings—such as social circles in social networks, industrial sectors in financial markets or biological modules—thereby offering explanatory and predictive insights into system behavior.  

However, beyond their descriptive utility, a central question remains: are the identified \textit{communities} statistically significant or are they artifacts of random connectivity? Even random networks can exhibit modular partitions with deceptively high modularity values \cite{hu2010measuring}, underscoring the need for rigorous statistical validation. A foundational framework by \cite{lancichinetti2010statistical} addressed this through a null-model approach based on the configuration model, comparing a community’s quality function \( q(C) \) against its expected value under a randomized null distribution \( q_{\text{null}} \). The corresponding tail probability,
\[
p = P\big(q_{\text{null}} \ge q(C)\big),
\]
quantifies the likelihood of observing such cohesive subgraphs by chance. This principle underpins the Order Statistics Local Optimization Method (OSLOM) algorithm \cite{lancichinetti2011finding}, which assigns significance scores to individual \textit{communities} rather than assuming all detected modules are meaningful.  

Building on this inferential foundation, subsequent studies extended statistical testing to diverse network settings. \cite{zhang2017hypothesis} formalized a hypothesis-testing framework under the degree-corrected stochastic block model (SBM), enabling rejection of the null hypothesis of no modular structure. \cite{kojaku2018core, kojaku2018generalised} advanced these ideas with a generalized statistical test that evaluates the significance of individual \textit{communities} independent of the detection algorithm, correcting for biases due to community size. Further developments by \cite{palowitch2018significance} introduced the Continuous Configuration Model Extraction (CCME) and False-Positive Optimized Community Significance (FOCS) methods for weighted networks, while \cite{yanchenko2024generalized} proposed a model-agnostic hypothesis test to assess the overall presence of significant modular organization. Collectively, these works transform community detection from a purely algorithmic pursuit into a problem of statistical inference, grounding it in probabilistic reasoning. Importantly, the notion of significance pertains to assortative density—connectivity exceeding random expectation—distinguishing it conceptually from CP organization, where structural prominence arises from hierarchical dominance rather than modular cohesion.

\section{Comparison Between CP and \textit{Community Structure}}

This section presents a comparative exploration of the CP and \textit{community structure}, emphasizing their sectoral representation, implications for portfolio optimization and relevance within the framework of hyperbolic network models. The analysis highlights how these mesoscale organizations complement and contrast each other in capturing the structural and functional heterogeneity of financial and complex systems. The subsections include: (i). Sectoral representation in CP and \textit{community structure},  (ii). Portfolio optimization using CP and \textit{community structure}, and (iii). CP and \textit{community structure} in the hyperbolic network model.

\subsection{Sectoral Representation in CP and \textit{Community Structure}}

In the context of financial networks, these structures capture different aspects of market organization. Community detection often aligns with sectoral classifications in the stock market, successfully clustering stocks that belong to the same industrial or economic sector. This indicates that stocks within the same sector tend to be more strongly correlated with each other than with those in other sectors, resulting in clear \textit{community structure}. On the other hand, \textit{\textit{CP structure}s} reflect hierarchical organization based on influence and connectivity. The core typically consists of large-cap or systemically important stocks—often from sectors such as finance that exert significant influence over the rest of the market. These core stocks are highly interconnected and maintain strong links with the periphery, which comprises less influential stocks with fewer interconnections. As a result, the CP decomposition provides insights into market dominance and influence dynamics that are not captured by \textit{community structure} alone.


RMT offers a powerful statistical framework for analyzing the spectral properties of correlation matrices in financial systems. When applied to stock return correlation matrices, RMT allows us to decompose the matrix into different eigenmodes—namely, the market mode, sector modes and noise. Empirical results show that the \textit{market mode}, associated with the largest eigenvalue, captures the global market dynamics and exhibits a pronounced \textit{CP structure}. This is because the most influential stocks, often forming the core drive the collective behavior of the market during periods of high volatility or systemic events. In contrast, the \textit{sector modes}, associated with intermediate eigenvalues, effectively reveal the underlying \textit{community structure} of the financial network. These modes isolate sector specific movements, highlighting groups of stocks that co-move due to common economic drivers or industrial factors. As a result, community detection applied to sector modes aligns well with known sector classifications. In conclusion, while both mesoscopic structures provide valuable insights, they capture different organizational layers of financial networks: the \textit{CP structure} is primarily associated with market wide systemic influence while the \textit{community structure} is more aligned with sectoral groupings and local interactions.

\subsection{Portfolio optimization using CP and \textit{Community Structure}}

Recent studies have demonstrated that CP and community detection methods provide valuable insights into the structural organization of complex systems. In particular, the modular configurations frequently observed in complex financial networks highlight the presence of both CP and \textit{community structure}s~\cite{zhao2021community, pawanesh2025exploring, ansari2025novel}. \textit{\textit{CP structure}s} in financial networks have gained attention for their ability to distinguish highly interconnected ``core'' assets from less connected ``peripheral'' assets. The core typically consists of stocks that are strongly correlated with each other and with the rest of the market, while the periphery contains assets with weaker connectivity, often capturing niche or idiosyncratic dynamics. In contrast, \textit{community structure} capture clusters of assets with high internal similarity, often reflecting sectoral, industrial or thematic groupings. Using these mesoscale organizations, researchers have proposed portfolio strategies to enhance diversification and manage systemic risk. Typically, \textit{\textit{CP structure}s} are extracted from correlation-based networks filtered using minimum spanning trees or \textit{Planar Maximally Filtered Graph}s and detected via algorithms such as the Rossa~\cite{rossa2013profiling} and the Rombach~\cite{rombach2017core} or via centrality-based approaches~\cite{pozzi2013spread, li2019portfolio, sharma2019mutual} whereas \textit{community structures} are identified using the the Louvain~\cite{blondel2008fast} and the Infomap~\cite{rosvall2011multilevel} algorithms. Below, we highlight several recent studies in which researchers have used these mesoscale structures to develop investment strategies aimed at constructing optimal portfolios.

In 2014, Pozzi \textit{et al.}~\cite{pozzi2013spread} conducted one of the seminal studies in the context of portfolio optimization by introducing a measure to identify \textit{core} and \textit{peripheral vertices} in financial networks. This \emph{hybrid measure} combines multiple centrality metrics, including \textit{degree centrality}, \textit{betweenness centrality}, \textit{eccentricity}, \textit{closeness} and \textit{eigenvector centrality}. Vertices with high hybrid scores are classified as \textit{peripheral} whereas those with low scores are considered part of the \textit{core}. The analysis is performed on networks constructed using the \textit{Planar Maximally Filtered Graph}~\cite{tumminello2005tool} derived from pairwise exponentially weighted correlations~\cite{pozzi2012exponential} in the American Stock Exchange market in the period from 1981 to 2010. A key finding of this study is that portfolios composed of peripheral stocks identified via the hybrid measure consistently outperform those composed of core stocks. The authors evaluate portfolio performance using both uniform weighting and the \textit{Markowitz mean-variance optimization} framework.

In 2021, Zhao \textit{et al.}~\cite{zhao2021community} used community insight in portfolio optimization and risk management. In this article, the authors constructed correlation-based networks in stock markets. They filtered them using the filtering approach, \textit{PMFG} and the Infomap algorithm to detect the \textit{community structure} and analyze the US, UK and Chinese stock markets. Next, they use this community information to construct portfolios with stocks from different \textit{communities} (inter-portfolios) and stocks from the same \textit{communities} (intra-portfolios). It has been shown that the intercommunity portfolios have lower risk and higher return in the US, UK and Chinese stock markets. Therefore, the risk diversification of the inter-community portfolios is much stronger than that of the intra-community portfolios. They have tested the portfolio size from 5 to 60 and have consistently obtained results. The good performance of these portfolios indicates the usefulness of \textit{community structure} in portfolio optimization. 

In 2025, Ansari \textit{et al.}~\cite{ansari2025novel} constructed correlation-based financial networks using both daily and high-frequency data. The raw correlation matrices were filtered using the \textit{PMFG} method to extract the most informative and economically meaningful connections. The \textit{CP structure} was then identified using continuous formulations, specifically the models proposed by Rossa \textit{et al.}~\cite{rossa2013profiling} and Rombach \textit{et al.}~\cite{rombach2017core}. Based on the resulting CP profiles, two distinct portfolio classes were constructed: one comprising core stocks and the other peripheral stocks. Portfolios were optimized under multiple weighting frameworks—uniform, risk parity \& Markowitz mean–variance optimization and benchmarked against both the Pozzi \textit{et al.}~\cite{pozzi2013spread} strategy and the market portfolio. Empirical analysis revealed that portfolios of peripheral stocks consistently outperformed their core counterparts across multiple evaluation metrics. Furthermore, peripheral-based portfolios not only exceeded the performance of the Pozzi benchmark but also exhibited comparable or superior risk-adjusted returns relative to the market portfolio as measured by the Sharpe ratio, average return and volatility (standard deviation). These findings highlight the underappreciated role of peripheral nodes, which, despite their weaker interconnections, contribute to enhanced diversification and reduced systemic risk exposure, making them valuable components for constructing robust and resilient portfolios.

In 2025, Pawanesh \textit{et al.}~\cite{pawanesh2025exploring} examined how community information can be leveraged in portfolio management to reduce risk and optimise returns. They first constructed financial networks using correlation based methods and then applied RMT to denoise the empirical correlation matrix by decomposing it into three components: the market mode (associated with the largest eigenvalue), the sector mode (capturing few significant eigenvalues) and the random mode (representing the remaining eigenvalues). The PMFG was subsequently applied to each filtered correlation matrix to remove insignificant links. \textit{community structure}s were then identified using the Louvain algorithm. Their findings showed that portfolios composed of inter-community stocks, constructed using the full correlation matrix, outperform those based solely on the sector mode. Furthermore, such portfolios also outperform market portfolios. This study demonstrates the effectiveness of full and RMT-filtered correlation matrices in improving our understanding of financial markets, particularly through the lens of \textit{community structure}.


\subsection{CP and \textit{Community Structure} in the hyperbolic networks model}

In the past, researchers have analyzed real-world networks and identified several characteristic properties of complex networks, including small-worldness~\cite{milgram1967small}, a relatively high clustering coefficient~\cite{watts1998collective}, heterogeneous degree distributions~\cite{faloutsos1999power}, \textit{community structure}~\cite{fortunato2010community, fortunato2016community} and the presence of \textit{CP structure}~\cite{holme2005core}. The interplay between nodes and edges in such networks reveals significant patterns and structural organization. Among the most intriguing models that capture these patterns, particularly based on node degree and similarity, are hyperbolic network models, such as the Popularity-Similarity Optimization (PSO) model~\cite{papadopoulos2012popularity} and the $\mathbb{S}^{1}/\mathbb{H}^{2}$ model~\cite{serrano2008self, garcia2019mercator}.

In 2021, Bianka \textit{et al.}~\cite{kovacs2021inherent} investigated the hidden \textit{community structure} of these hyperbolic network models and discussed them for a wide range of parameter settings. The authors generated the networks with a range of parameters such as popularity fading $\beta$ and temperature $T$ (average clustering coefficient) that varied in the plane $(0,1]\times[0,1)$ and the analogous parameters $(1/(\gamma-1),1/\alpha)$, where $\gamma$ is the power law coefficient, $\alpha$ (average clustering coefficient) in the plane $(0,1)\times(0,1)$ for the \textit{PSO model} and $\mathbb{S}^{1}/\mathbb{H}^{2}$, respectively. Then, they studied the \textit{community structure} using various community detection algorithms. They claim that as the parameter settings go to the origin in both hyperbolic models, they yield the best \textit{community structure}. In addition, they analyzed the \textit{community structure} as a function of the number of network vertices. They claimed that the \textit{community structure} improves for almost all parameter settings for both models. For a detailed study, we refer the reader to the original study~\cite{kovacs2021inherent}.

In 2025, Ansari \textit{et al.}~\cite{ansari2025uncovering} uncovered the \textit{CP structure} in hyperbolic network models, specifically the \textit{PSO} and $\mathbb{S}^{1}/\mathbb{H}^{2}$ models, across a wide range of parameter settings. This work thoroughly investigated these structures using well-established \textit{CP structure} algorithms~\cite{rossa2013profiling, rombach2017core} using the CP centralization index. Furthermore, this work validates the findings through rigorous statistical testing to assess the significance of the observed \textit{\textit{CP structure}s} in these hyperbolic network models. This study focuses on extending the modeling capabilities of hyperbolic models, particularly within the context of \textit{\textit{CP structure}s}. Although these models, as discussed in the literature~\cite{garcia2019mercator,papadopoulos2014network}, capture essential network properties and their applicability to \textit{\textit{CP structure}s} remains underexplored. This study involves extensive simulations, parameter space exploration and advanced CP detection algorithms to identify CP regions. The results have implications for modeling real world networks with CP organizations and contribute to the broader understanding of network science. This research represents a concerted effort to advance our knowledge of the structures of the CP within hyperbolic network models, particularly in the context of \textit{PSO} and the $\mathbb{S}^{1}/\mathbb{H}^{2}$ model.

\section{Differentiation Between CP and \textit{Community Structure}}

In complex networks, the CP and \textit{community structure} represent two distinct forms of mesoscale organization~\cite{borgatti2000models,newman2004finding,newman2006finding,rombach2017core}. Although both the CP and \textit{community structure} reveal patterns of organization within the networks, they capture fundamentally different aspects of connectivity. The CP organization characterizes a network through a hierarchical arrangement in which a densely interconnected core coexists with a sparsely connected periphery that primarily links to the core but rarely to other peripheral nodes~\cite{borgatti2000models}. In contrast, a \textit{community structure} captures the tendency of nodes to form densely connected groups (\textit{communities} or modules) that are only sparsely linked to nodes in other groups~\cite{newman2004finding}.

In 2018, Kojaku \textit{et al.}~\cite{kojaku2018core} compared the core--periphery pairs identified by the KM--config model with the \textit{community structure} in various empirical networks. \textit{communities} were determined through modularity maximization using the Louvain algorithm~\cite{blondel2008fast}. The algorithm was executed ten times and the node partition yielding the highest modularity value was selected. Kojaku \textit{et al.} reported the modularity values for both the partitions obtained by the Louvain algorithm and those derived from the KM--config method, including the insignificant core--periphery pairs. They observed that the modularity values of the core--periphery partitions were close to those obtained via modularity maximization for most empirical networks. This finding suggests that the detected core--periphery pairs may exhibit structural similarities to \textit{communities}. This result raises an important question: Is a CP pair essentially a community in the traditional sense? If so, does the KM--config algorithm effectively classify the nodes within each community into a core and a periphery based on the composition of intra- and inter-community edges associated with each node?

To examine whether core--periphery pairs correspond to traditional \textit{communities}, they analyze the role of each node using the cartographic representation of networks~\cite{guimera2005functional}, where each node $i$ is characterized by the standardized within-pair degree $z_i$ and participation coefficient $p_i$. The within-pair degree is defined as $z_i = (\tilde{d}_{i,c_i} - \langle \tilde{d}_{j,c_i} \rangle_j)/s_{c_i}$, where $\tilde{d}_{i,c_i}$ is the number of neighbours of node $i$ in its core--periphery pair $c_i$, $\langle \tilde{d}_{j,c_i} \rangle_j$ is the average internal degree over all nodes in that pair and $s_{c_i}$ is the unbiased standard deviation of $\tilde{d}_{j,c_i}$. The participation coefficient is $p_i = 1 - \sum_{c=1}^{C} (\tilde{d}_{i,c}/d_i)^2$, where $\tilde{d}_{i,c}$ is the number of neighbours of node $i$ in pair $c$ and $d_i = \sum_{c=1}^C \tilde{d}_{i,c}$ is the total degree. High $z_i$ indicates core-like nodes with many intra-pair connections, while $p_i$ near 0 implies connections mainly within a single pair and $p_i$ near 1 implies connections distributed across many pairs. In the $(z_i, p_i)$ plane, the core nodes typically have high $z_i$ and low $p_i$, while the peripheral nodes have lower $z_i$; on the contrary, community nodes show a high intra-community density (high $z_i$) and generally low $p_i$, reflecting the membership of the cohesive module. Therefore, while community detection and CP analysis may produce similar modular partitions at a macroscopic level, their underlying node role characterizations differ substantially.


\noindent
In 2018, Yang \textit{et al.}~\cite{yang2018structural} proposed that \textit{community structure} may be equivalent to \textit{\textit{CP structure}s} in networks characterized by unequal influence and asymmetric interactions, such as social media. They note that some community detection algorithms implicitly assume a leader-follower relationship, viewing a community as a group of followers organized around potential leaders. In these approaches, a community closely resembles a \textit{CP structure}, where leaders correspond to core actors and followers occupy peripheral positions within the community. Therefore, they propose that \textit{communities} identified from social media using community detection methods inherently exhibit a \textit{CP structure}. Their experimental results using Twitter data provide empirical evidence supporting this claim.

In 2022, Wedell \textit{et al.}~\cite{wedell2022center} investigated whether research \textit{communities}, as represented in citation networks, exhibit a \textit{CP structure} within their community organization. The authors designed a four-stage analytic pipeline to identify publication \textit{communities} that display a CP substructure. Their main goal was to determine whether the graphical structure of such \textit{communities}, based on citation patterns, reflects the social structure of scientific collaborations, such as groups of coauthors. Applying this method to a large citation network in the field of exosomes and extracellular vesicles, they found that many publication \textit{communities} indeed show a CP pattern, with a dense, highly connected ``core'' of influential publications surrounded by a ``periphery'' of less connected papers. This result supports their hypothesis that \textit{communities} with CP organization exist in scientific research domains. The study also highlights that their pipeline can successfully detect such structures, although it was tested on a single field. They note that future research should examine whether this pattern generalizes to other disciplines and whether alternative CP models \cite{borgatti2000models,boyd2010computing,rossa2013profiling,rombach2017core} might provide deeper insights into the organization of research \textit{communities}.



\section{Open Challenges}

Mesoscale structures lie at the heart of complex network analysis, offering profound insights into how real-world systems are organized and function. While community detection has reached a relatively mature stage, with well-established algorithms and theoretical foundations, the study of \textit{\textit{CP structure}s} is still evolving, driven by recent methodological advances and expanding domains of application. Understanding how these two mesoscale patterns coexist and interact remains an open and fascinating challenge, with implications for both theory and practice. Addressing these issues can lead to new frameworks for modeling hierarchical and modular organization across diverse fields. Below, we outline several promising directions and unresolved questions concerning the exploration of CP and \textit{community structure} and their applications in various real-world systems.

\begin{itemize}

\item The \textit{community structure} has been extensively studied in temporal~\cite{peixoto2017modelling,gauvin2014detecting,he2015fast,linhares2019scalable} and multi-layer networks~\cite{jeub2017local,pramanik2017discovering,khawaja2021uncovering,wilson2017community}, achieving remarkable performance. While the \textit{CP structure} has received some attention in multi-layer networks~\cite{bergermann2025core,beranek2023emergence}, it remains largely unexplored in temporal networks. Given that several studies have demonstrated the comparable importance of CP and \textit{community structure}, future research should aim to extend CP analysis to temporal and multilayer network frameworks.

\item RMT has played a pivotal role in addressing real-world problems, particularly in filtering noise from empirical correlation matrices in complex systems such as financial markets~\cite{pan2007collective, pawanesh2025exploring, jiang2014structure}. For instance, Ref.~\cite{pawanesh2025exploring} analyzed the CP and \textit{community structure} in daily stock price networks using RMT, highlighting its utility in revealing underlying mesoscale organization. However, several open challenges remain. One promising direction is to incorporate high-frequency (intraday) financial data, which can capture short-term fluctuations and transient correlations often obscured in daily analyses. Such an extension could reveal how market structures evolve across multiple time scales, providing deeper insights into the dynamic formation and temporal stability of CP relationships. 

Another critical avenue is the cross-market generalization of RMT-based filtering techniques. Investigating their robustness across different asset classes, market regimes and geographical regions could help determine the universality or market-specific nature of RMT-derived structural patterns. Beyond financial systems, the methodological framework of RMT-based correlation filtering could be adapted to other complex domains, including social, biological and transportation networks, where high-dimensional noisy data often obscure latent organizational patterns. Addressing these challenges would advance both the theoretical foundation and practical applicability of RMT, paving the way toward a multiscale, cross domain theory of noise-filtered network structures.

\item Recent advances in theory have significantly improved the understanding of complex systems, particularly the \textit{\textit{CP structure}s}. Recent works~\cite{kovacs2021inherent, ansari2025uncovering} highlight that traditional hyperbolic models have provided more profound insights about the CP  and \textit{community structure}; on the other hand, extending this study to generalized hyperbolic network models, such as the \textit{d-PSO}~\cite{kovacs2022generalised} and $\mathbb{S}^{D}/\mathbb{H}^{D+1}$~\cite{jankowski2023d, budel2024random} models, could uncover more profound insights into network organization and underlying mechanisms and hence could be a promising area for future research, particularly to investigate the relationship between multidimensionality and the \textit{CP structure}. Furthermore, investigating these generalized models may provide even more insight into their suitability for capturing complex real-world systems, broadening our understanding of the interaction between hyperbolic geometry and \textit{\textit{CP structure}s}.

\item Many studies contribute to the growing literature on network-based portfolio optimization by demonstrating that peripheral stocks, identified through network centrality in financial networks, can yield superior risk-adjusted returns compared to conventional strategies~\cite{pozzi2013spread, ansari2025novel, li2019portfolio}. For example, Ref.~\cite{ansari2025novel} underscores the importance of mesoscale network structures in designing alternative investment strategies beyond traditional centrality-based~\cite{pozzi2013spread, sharma2019mutual} or correlation-based approaches~\cite{li2019portfolio,ansari2025novel}. Despite these promising findings, several open problems remain to be explored. Future research could incorporate comparative analyses with advanced portfolio construction frameworks such as risk parity, multi-factor or machine learning–based optimization methods, to position the CP approach within the broader landscape of quantitative finance. Another promising direction lies in extending the framework to regime-switching models or volatility-based segmentation, enabling a more nuanced understanding of market dynamics and structural transitions over time.

\item Recently, the detection of CP~\cite{tudisco2023core,papachristou2022core} and community~\cite{yuan2022testing,ruggeri2024framework} structures in hypergraphs has gained significant attention. Moreover, integrating hypergraph-based network representations presents a particularly compelling avenue for future exploration. Hypergraphs can capture higher-order dependencies among multiple stocks beyond pairwise correlations, potentially uncovering richer relational and structural patterns that influence portfolio performance. Developing such extensions would significantly enhance both the theoretical foundations and practical relevance of network-based portfolio strategies, paving the way toward a comprehensive and robust framework for data-driven financial decision making.
\end{itemize}
As the above open questions illustrate, despite decades of active research, the study of CP and \textit{community structures} remains a vibrant and promising field. The authors encourage fellow researchers to continue exploring these fascinating phenomena and contribute to advancing our understanding of complex networks.

\section*{Acknowledgments}

The authors thank Dr. Qazi J Azhad for their valuable suggestions.

\bibliographystyle{unsrtnat}
\bibliography{bib}

\begin{thebibliography}{157}
\providecommand{\natexlab}[1]{#1}
\providecommand{\url}[1]{\texttt{#1}}
\expandafter\ifx\csname urlstyle\endcsname\relax
  \providecommand{\doi}[1]{doi: #1}\else
  \providecommand{\doi}{doi: \begingroup \urlstyle{rm}\Url}\fi

\bibitem[Newman(2018)]{newman2018networks}
Mark Newman.
\newblock \emph{Networks}.
\newblock Oxford university press, 2018.

\bibitem[Gulati et~al.(2000)Gulati, Nohria, and Zaheer]{gulati2000strategic}
Ranjay Gulati, Nitin Nohria, and Akbar Zaheer.
\newblock Strategic networks.
\newblock \emph{Strategic management journal}, 21\penalty0 (3):\penalty0 203--215, 2000.

\bibitem[de~Silva and Stumpf(2005)]{de2005complex}
Eric de~Silva and Michael~PH Stumpf.
\newblock Complex networks and simple models in biology.
\newblock \emph{Journal of the Royal Society Interface}, 2\penalty0 (5):\penalty0 419--430, 2005.

\bibitem[Costa et~al.(2008)Costa, Rodrigues, and Cristino]{costa2008complex}
Luciano da~F Costa, Francisco~A Rodrigues, and Alexandre~S Cristino.
\newblock Complex networks: the key to systems biology.
\newblock \emph{Genetics and Molecular Biology}, 31:\penalty0 591--601, 2008.

\bibitem[Albert and Barab{\'a}si(2002)]{albert2002statistical}
R{\'e}ka Albert and Albert-L{\'a}szl{\'o} Barab{\'a}si.
\newblock Statistical mechanics of complex networks.
\newblock \emph{Reviews of modern physics}, 74\penalty0 (1):\penalty0 47, 2002.

\bibitem[Sun and Wandelt(2021)]{sun2021robustness}
Xiaoqian Sun and Sebastian Wandelt.
\newblock Robustness of air transportation as complex networks: Systematic review of 15 years of research and outlook into the future.
\newblock \emph{Sustainability}, 13\penalty0 (11):\penalty0 6446, 2021.

\bibitem[Newman(2003{\natexlab{a}})]{newman2003structure}
Mark~EJ Newman.
\newblock The structure and function of complex networks.
\newblock \emph{SIAM review}, 45\penalty0 (2):\penalty0 167--256, 2003{\natexlab{a}}.

\bibitem[Boccaletti et~al.(2006)Boccaletti, Latora, Moreno, Chavez, and Hwang]{boccaletti2006complex}
Stefano Boccaletti, Vito Latora, Yamir Moreno, Martin Chavez, and D-U Hwang.
\newblock Complex networks: Structure and dynamics.
\newblock \emph{Physics reports}, 424\penalty0 (4-5):\penalty0 175--308, 2006.

\bibitem[Lin and Ban(2013)]{lin2013complex}
Jingyi Lin and Yifang Ban.
\newblock Complex network topology of transportation systems.
\newblock \emph{Transport reviews}, 33\penalty0 (6):\penalty0 658--685, 2013.

\bibitem[H{\'a}znagy et~al.(2015)H{\'a}znagy, Fi, London, and Nemeth]{haznagy2015complex}
Andor H{\'a}znagy, Istv{\'a}n Fi, Andr{\'a}s London, and Tam{\'a}s Nemeth.
\newblock Complex network analysis of public transportation networks: A comprehensive study.
\newblock In \emph{2015 International Conference on Models and Technologies for Intelligent Transportation Systems (MT-ITS)}, pages 371--378. IEEE, 2015.

\bibitem[Mata(2020)]{mata2020complex}
Ang{\'e}lica Sousa~da Mata.
\newblock Complex networks: a mini-review.
\newblock \emph{Brazilian Journal of Physics}, 50\penalty0 (5):\penalty0 658--672, 2020.

\bibitem[Spirin and Mirny(2003)]{spirin2003protein}
Victor Spirin and Leonid~A Mirny.
\newblock Protein complexes and functional modules in molecular networks.
\newblock \emph{Proceedings of the national Academy of sciences}, 100\penalty0 (21):\penalty0 12123--12128, 2003.

\bibitem[Tornow and Mewes(2003)]{tornow2003functional}
Sabine Tornow and HW~Mewes.
\newblock Functional modules by relating protein interaction networks and gene expression.
\newblock \emph{Nucleic acids research}, 31\penalty0 (21):\penalty0 6283--6289, 2003.

\bibitem[Chen and Yuan(2006)]{chen2006detecting}
Jingchun Chen and Bo~Yuan.
\newblock Detecting functional modules in the yeast protein--protein interaction network.
\newblock \emph{Bioinformatics}, 22\penalty0 (18):\penalty0 2283--2290, 2006.

\bibitem[Girvan and Newman(2002)]{girvan2002community}
Michelle Girvan and Mark~EJ Newman.
\newblock Community structure in social and biological networks.
\newblock \emph{Proceedings of the national academy of sciences}, 99\penalty0 (12):\penalty0 7821--7826, 2002.

\bibitem[Newman and Girvan(2004)]{newman2004finding}
Mark~EJ Newman and Michelle Girvan.
\newblock Finding and evaluating community structure in networks.
\newblock \emph{Physical review E}, 69\penalty0 (2):\penalty0 026113, 2004.

\bibitem[Bargigli and Gallegati(2013)]{bargigli2013finding}
Leonardo Bargigli and Mauro Gallegati.
\newblock Finding communities in credit networks.
\newblock \emph{Economics}, 7\penalty0 (1):\penalty0 20130017, 2013.

\bibitem[Chan-Lau(2018)]{chan2018systemic}
Jorge~A Chan-Lau.
\newblock Systemic centrality and systemic communities in financial networks.
\newblock \emph{Quantitative Finance and Economics}, 2\penalty0 (2):\penalty0 468--496, 2018.

\bibitem[Zhao et~al.(2021)Zhao, Wang, Wang, Stanley, and Chen]{zhao2021community}
Longfeng Zhao, Chao Wang, Gang-Jin Wang, H~Eugene Stanley, and Lin Chen.
\newblock Community detection and portfolio optimization.
\newblock \emph{arXiv preprint arXiv:2112.13383}, 2021.

\bibitem[Huang and Chen(2021)]{huang2021community}
Yajing Huang and Feng Chen.
\newblock Community structure and systemic risk of bank correlation networks based on the us financial crisis in 2008.
\newblock \emph{Algorithms}, 14\penalty0 (6):\penalty0 162, 2021.

\bibitem[Pawanesh et~al.(2025{\natexlab{a}})Pawanesh, Ansari, and Sahni]{pawanesh2025exploring}
Pawanesh Pawanesh, Imran Ansari, and Niteesh Sahni.
\newblock Exploring the core--periphery and community structure in the financial networks through random matrix theory.
\newblock \emph{Physica A: Statistical Mechanics and its Applications}, 661:\penalty0 130403, 2025{\natexlab{a}}.

\bibitem[Nie and Song(2018)]{nie2018constructing}
Chun-Xiao Nie and Fu-Tie Song.
\newblock Constructing financial network based on pmfg and threshold method.
\newblock \emph{Physica A: Statistical Mechanics and its Applications}, 495:\penalty0 104--113, 2018.

\bibitem[Pawanesh et~al.(2025{\natexlab{b}})Pawanesh, Sharma, and Sahni]{pawanesh2025exploiting}
Pawanesh, Charu Sharma, and Niteesh Sahni.
\newblock Exploiting the geometry of heterogeneous networks: a case study of the indian stock market: P. yadav et al.
\newblock \emph{Soft Computing}, pages 1--18, 2025{\natexlab{b}}.

\bibitem[Tun{\c{c}} and Verma(2015{\natexlab{a}})]{tuncc2015unifying}
Birkan Tun{\c{c}} and Ragini Verma.
\newblock Unifying inference of meso-scale structures in networks.
\newblock \emph{PloS one}, 10\penalty0 (11):\penalty0 e0143133, 2015{\natexlab{a}}.

\bibitem[Rombach et~al.(2017)Rombach, Porter, Fowler, and Mucha]{rombach2017core}
Puck Rombach, Mason~A Porter, James~H Fowler, and Peter~J Mucha.
\newblock Core-periphery structure in networks (revisited).
\newblock \emph{SIAM review}, 59\penalty0 (3):\penalty0 619--646, 2017.

\bibitem[Borgatti and Everett(1999)]{borgatti2000models}
Stephen~P Borgatti and Martin~G Everett.
\newblock Models of corerperiphery structures, 1999.
\newblock URL \url{www.elsevier.comrlocatersocnet}.

\bibitem[Holme(2005)]{holme2005core}
Petter Holme.
\newblock Core-periphery organization of complex networks.
\newblock \emph{Physical Review E—Statistical, Nonlinear, and Soft Matter Physics}, 72\penalty0 (4):\penalty0 046111, 2005.

\bibitem[Boyd et~al.(2010{\natexlab{a}})Boyd, Fitzgerald, Mahutga, and Smith]{boyd2010computing}
John~P Boyd, William~J Fitzgerald, Matthew~C Mahutga, and David~A Smith.
\newblock Computing continuous core/periphery structures for social relations data with minres/svd.
\newblock \emph{Social Networks}, 32\penalty0 (2):\penalty0 125--137, 2010{\natexlab{a}}.

\bibitem[Rossa et~al.(2013)Rossa, Dercole, and Piccardi]{rossa2013profiling}
Fabio~Della Rossa, Fabio Dercole, and Carlo Piccardi.
\newblock Profiling core-periphery network structure by random walkers.
\newblock \emph{Scientific reports}, 3\penalty0 (1):\penalty0 1467, 2013.

\bibitem[Kojaku and Masuda(2018{\natexlab{a}})]{kojaku2018core}
Sadamori Kojaku and Naoki Masuda.
\newblock Core-periphery structure requires something else in the network.
\newblock \emph{New Journal of physics}, 20\penalty0 (4):\penalty0 043012, 2018{\natexlab{a}}.

\bibitem[Kojaku and Masuda(2017)]{kojaku2017finding}
Sadamori Kojaku and Naoki Masuda.
\newblock Finding multiple core-periphery pairs in networks.
\newblock \emph{Physical Review E}, 96\penalty0 (5):\penalty0 052313, 2017.

\bibitem[Newman(2004{\natexlab{a}})]{Newman2004}
M.~E. Newman.
\newblock Fast algorithm for detecting community structure in networks.
\newblock \emph{Physical Review E}, 69:\penalty0 066133, 2004{\natexlab{a}}.
\newblock \doi{10.1103/PhysRevE.69.066133}.

\bibitem[Newman(2006{\natexlab{a}})]{newman2006finding}
Mark~EJ Newman.
\newblock Finding community structure in networks using the eigenvectors of matrices.
\newblock \emph{Physical Review E—Statistical, Nonlinear, and Soft Matter Physics}, 74\penalty0 (3):\penalty0 036104, 2006{\natexlab{a}}.

\bibitem[Blondel et~al.(2008)Blondel, Guillaume, Lambiotte, and Lefebvre]{blondel2008fast}
Vincent~D Blondel, Jean-Loup Guillaume, Renaud Lambiotte, and Etienne Lefebvre.
\newblock Fast unfolding of communities in large networks.
\newblock \emph{Journal of statistical mechanics: theory and experiment}, 2008\penalty0 (10):\penalty0 P10008, 2008.

\bibitem[Raghavan et~al.(2007)Raghavan, Albert, and Kumara]{raghavan2007near}
Usha~Nandini Raghavan, R{\'e}ka Albert, and Soundar Kumara.
\newblock Near linear time algorithm to detect community structures in large-scale networks.
\newblock \emph{Physical Review E—Statistical, Nonlinear, and Soft Matter Physics}, 76\penalty0 (3):\penalty0 036106, 2007.

\bibitem[Lou et~al.(2013)Lou, Li, and Zhao]{lou2013detecting}
Hao Lou, Shenghong Li, and Yuxin Zhao.
\newblock Detecting community structure using label propagation with weighted coherent neighborhood propinquity.
\newblock \emph{Physica A: Statistical Mechanics and its Applications}, 392\penalty0 (14):\penalty0 3095--3105, 2013.

\bibitem[Zhang et~al.(2017)Zhang, Ren, Song, Jia, and Zhang]{zhang2017label}
Xian-Kun Zhang, Jing Ren, Chen Song, Jia Jia, and Qian Zhang.
\newblock Label propagation algorithm for community detection based on node importance and label influence.
\newblock \emph{Physics Letters A}, 381\penalty0 (33):\penalty0 2691--2698, 2017.

\bibitem[Berahmand and Bouyer(2018)]{berahmand2018lp}
Kamal Berahmand and Asgarali Bouyer.
\newblock Lp-lpa: A link influence-based label propagation algorithm for discovering community structures in networks.
\newblock \emph{International Journal of Modern Physics B}, 32\penalty0 (06):\penalty0 1850062, 2018.

\bibitem[Shen et~al.(2010)Shen, Cheng, and Fang]{shen2010covariance}
Hua-Wei Shen, Xue-Qi Cheng, and Bin-Xing Fang.
\newblock Covariance, correlation matrix, and the multiscale community structure of networks.
\newblock \emph{Physical Review E—Statistical, Nonlinear, and Soft Matter Physics}, 82\penalty0 (1):\penalty0 016114, 2010.

\bibitem[Xie et~al.(2009)Xie, Ji, Zhang, and Huang]{xie2009detection}
Fuding Xie, Min Ji, Yong Zhang, and Dan Huang.
\newblock The detection of community structure in network via an improved spectral method.
\newblock \emph{Physica A: Statistical Mechanics and its Applications}, 388\penalty0 (15-16):\penalty0 3268--3272, 2009.

\bibitem[Ansari et~al.(2024)Ansari, Azhad, and Sahni]{ansari2024identifying}
Imran Ansari, Qazi~J Azhad, and Niteesh Sahni.
\newblock Identifying core-periphery structures in networks via artificial ants.
\newblock \emph{arXiv preprint arXiv:2411.11900}, 2024.

\bibitem[Mohamed et~al.(2019)Mohamed, Agouti, Tikniouine, and El~Adnani]{mohamed2019comprehensive}
El-Moussaoui Mohamed, Tarik Agouti, Abdessadek Tikniouine, and Mohamed El~Adnani.
\newblock A comprehensive literature review on community detection: Approaches and applications.
\newblock \emph{Procedia Computer Science}, 151:\penalty0 295--302, 2019.

\bibitem[Jin et~al.(2021)Jin, Yu, Jiao, Pan, He, Wu, Yu, and Zhang]{jin2021survey}
Di~Jin, Zhizhi Yu, Pengfei Jiao, Shirui Pan, Dongxiao He, Jia Wu, Philip~S Yu, and Weixiong Zhang.
\newblock A survey of community detection approaches: From statistical modeling to deep learning.
\newblock \emph{IEEE Transactions on Knowledge and Data Engineering}, 35\penalty0 (2):\penalty0 1149--1170, 2021.

\bibitem[Dey et~al.(2022)Dey, Tian, and Gel]{dey2022community}
Asim~K Dey, Yahui Tian, and Yulia~R Gel.
\newblock Community detection in complex networks: From statistical foundations to data science applications.
\newblock \emph{Wiley Interdisciplinary Reviews: Computational Statistics}, 14\penalty0 (2):\penalty0 e1566, 2022.

\bibitem[Nooribakhsh et~al.(2024)Nooribakhsh, Fern{\'a}ndez-Diego, Gonz{\'a}lez-Ladr{\'o}n-De-Guevara, and Mollamotalebi]{nooribakhsh2024community}
Mahsa Nooribakhsh, Marta Fern{\'a}ndez-Diego, Fernando Gonz{\'a}lez-Ladr{\'o}n-De-Guevara, and Mahdi Mollamotalebi.
\newblock Community detection in social networks using machine learning: a systematic mapping study.
\newblock \emph{Knowledge and Information Systems}, 66\penalty0 (12):\penalty0 7205--7259, 2024.

\bibitem[Izem et~al.(2024)Izem, Zaz, Younes, and El~Rharbi]{izem2024survey}
Oumaima Izem, Youssef Zaz, Ali Younes, and Nassime El~Rharbi.
\newblock A survey on community identification in dynamic network.
\newblock In \emph{E-Learning and Smart Engineering Systems (ELSES 2023)}, pages 335--351. Atlantis Press, 2024.

\bibitem[Yanchenko and Sengupta(2023)]{yanchenko2023core}
Eric Yanchenko and Srijan Sengupta.
\newblock Core-periphery structure in networks: A statistical exposition.
\newblock \emph{Statistic Surveys}, 17:\penalty0 42--74, 2023.

\bibitem[Csermely et~al.(2013)Csermely, London, Wu, and Uzzi]{csermely2013structure}
Peter Csermely, Andr{\'a}s London, Ling-Yun Wu, and Brian Uzzi.
\newblock Structure and dynamics of core/periphery networks.
\newblock \emph{Journal of Complex Networks}, 1\penalty0 (2):\penalty0 93--123, 2013.

\bibitem[Gallagher et~al.(2021)Gallagher, Young, and Welles]{gallagher2021clarified}
Ryan~J Gallagher, Jean-Gabriel Young, and Brooke~Foucault Welles.
\newblock A clarified typology of core-periphery structure in networks.
\newblock \emph{Science advances}, 7\penalty0 (12):\penalty0 eabc9800, 2021.

\bibitem[Watts and Strogatz(1998)]{watts1998collective}
Duncan~J Watts and Steven~H Strogatz.
\newblock Collective dynamics of ‘small-world’networks.
\newblock \emph{nature}, 393\penalty0 (6684):\penalty0 440--442, 1998.

\bibitem[Kov{\'a}cs and Palla(2021)]{kovacs2021inherent}
Bianka Kov{\'a}cs and Gergely Palla.
\newblock The inherent community structure of hyperbolic networks.
\newblock \emph{Scientific Reports}, 11\penalty0 (1):\penalty0 16050, 2021.

\bibitem[Coscia et~al.(2011)Coscia, Giannotti, and Pedreschi]{coscia2011classification}
Michele Coscia, Fosca Giannotti, and Dino Pedreschi.
\newblock A classification for community discovery methods in complex networks.
\newblock \emph{Statistical Analysis and Data Mining: The ASA Data Science Journal}, 4\penalty0 (5):\penalty0 512--546, 2011.

\bibitem[Yang et~al.(2018)Yang, Zhang, Shen, Ju, and Guo]{yang2018structural}
Jinfeng Yang, Min Zhang, Kathy~Ning Shen, Xiaofeng Ju, and Xitong Guo.
\newblock Structural correlation between communities and core-periphery structures in social networks: Evidence from twitter data.
\newblock \emph{Expert Systems with Applications}, 111:\penalty0 91--99, 2018.

\bibitem[Gamble et~al.(2016)Gamble, Chintakunta, Wilkerson, and Krim]{gamble2016node}
Jennifer Gamble, Harish Chintakunta, Adam Wilkerson, and Hamid Krim.
\newblock Node dominance: Revealing community and core-periphery structure in social networks.
\newblock \emph{IEEE Transactions on Signal and Information Processing over Networks}, 2\penalty0 (2):\penalty0 186--199, 2016.

\bibitem[De~Leo et~al.(2013)De~Leo, Santoboni, Cerina, Mureddu, Secchi, and Chessa]{de2013community}
Vincenzo De~Leo, Giovanni Santoboni, Federica Cerina, Mario Mureddu, Luca Secchi, and Alessandro Chessa.
\newblock Community core detection in transportation networks.
\newblock \emph{Physical Review E—Statistical, Nonlinear, and Soft Matter Physics}, 88\penalty0 (4):\penalty0 042810, 2013.

\bibitem[Lee et~al.(2014)Lee, Cucuringu, and Porter]{lee2014density}
Sang~Hoon Lee, Mihai Cucuringu, and Mason~A Porter.
\newblock Density-based and transport-based core-periphery structures in networks.
\newblock \emph{Physical Review E}, 89\penalty0 (3):\penalty0 032810, 2014.

\bibitem[Wilson et~al.(2016)Wilson, D~WILKINS, Lin, Lua, and Lichtarge]{wilson2016discovery}
Stephen~J Wilson, ANGELA D~WILKINS, Chih-Hsu Lin, Rhonald~C Lua, and Olivier Lichtarge.
\newblock Discovery of functional and disease pathways by community detection in protein-protein interaction networks.
\newblock In \emph{Pacific Symposium on Biocomputing. Pacific Symposium on Biocomputing}, volume~22, page 336, 2016.

\bibitem[Luo et~al.(2009)Luo, Li, Wan, and Scheuermann]{luo2009core}
Feng Luo, Bo~Li, Xiu-Feng Wan, and Richard~H Scheuermann.
\newblock Core and periphery structures in protein interaction networks.
\newblock \emph{BMC bioinformatics}, 10\penalty0 (Suppl 4):\penalty0 S8, 2009.

\bibitem[Newman(2004{\natexlab{b}})]{newman2004fast}
Mark~EJ Newman.
\newblock Fast algorithm for detecting community structure in networks.
\newblock \emph{Physical Review E—Statistical, Nonlinear, and Soft Matter Physics}, 69\penalty0 (6):\penalty0 066133, 2004{\natexlab{b}}.

\bibitem[Clauset et~al.(2004)Clauset, Newman, and Moore]{clauset2004finding}
Aaron Clauset, M.~E.~J. Newman, and Cristopher Moore.
\newblock Finding community structure in very large networks.
\newblock \emph{Physical Review E}, 70\penalty0 (6):\penalty0 066111, 2004.
\newblock \doi{10.1103/PhysRevE.70.066111}.

\bibitem[Wakita and Tsurumi(2007)]{wakita2007finding}
Ken Wakita and Toshiyuki Tsurumi.
\newblock Finding community structure in mega-scale social networks.
\newblock In \emph{Proceedings of the 16th international conference on World Wide Web}, pages 1275--1276, 2007.

\bibitem[Newman(2013)]{newman2013spectral}
Mark~EJ Newman.
\newblock Spectral methods for community detection and graph partitioning.
\newblock \emph{Physical Review E—Statistical, Nonlinear, and Soft Matter Physics}, 88\penalty0 (4):\penalty0 042822, 2013.

\bibitem[Zhang et~al.(2015)Zhang, Martin, and Newman]{zhang2015identification}
Xiao Zhang, Travis Martin, and Mark~EJ Newman.
\newblock Identification of core-periphery structure in networks.
\newblock \emph{Physical Review E}, 91\penalty0 (3):\penalty0 032803, 2015.

\bibitem[Peixoto(2019)]{peixoto2019bayesian}
Tiago~P Peixoto.
\newblock Bayesian stochastic blockmodeling.
\newblock \emph{Advances in network clustering and blockmodeling}, pages 289--332, 2019.

\bibitem[Cucuringu et~al.(2016)Cucuringu, Rombach, Lee, and Porter]{cucuringu2016detection}
Mihai Cucuringu, Puck Rombach, Sang~Hoon Lee, and Mason~A Porter.
\newblock Detection of core--periphery structure in networks using spectral methods and geodesic paths.
\newblock \emph{European Journal of Applied Mathematics}, 27\penalty0 (6):\penalty0 846--887, 2016.

\bibitem[Ma et~al.(2018)Ma, Xiang, Chen, Small, and Zhang]{ma2018detection}
Chuang Ma, Bing-Bing Xiang, Han-Shuang Chen, Michael Small, and Hai-Feng Zhang.
\newblock Detection of core-periphery structure in networks based on 3-tuple motifs.
\newblock \emph{Chaos: An Interdisciplinary Journal of Nonlinear Science}, 28\penalty0 (5), 2018.

\bibitem[Malliaros et~al.(2020)Malliaros, Giatsidis, Papadopoulos, and Vazirgiannis]{malliaros2020core}
Fragkiskos~D Malliaros, Christos Giatsidis, Apostolos~N Papadopoulos, and Michalis Vazirgiannis.
\newblock The core decomposition of networks: Theory, algorithms and applications.
\newblock \emph{The VLDB Journal}, 29\penalty0 (1):\penalty0 61--92, 2020.

\bibitem[Da~Silva et~al.(2008)Da~Silva, Ma, and Zeng]{da2008centrality}
Marcio~Rosa Da~Silva, Hongwu Ma, and An-Ping Zeng.
\newblock Centrality, network capacity, and modularity as parameters to analyze the core-periphery structure in metabolic networks.
\newblock \emph{Proceedings of the IEEE}, 96\penalty0 (8):\penalty0 1411--1420, 2008.

\bibitem[Meyer(2023)]{meyer2023matrix}
Carl~D Meyer.
\newblock \emph{Matrix analysis and applied linear algebra}.
\newblock SIAM, 2023.

\bibitem[Jia and Benson(2019)]{jia2019random}
Junteng Jia and Austin~R Benson.
\newblock Random spatial network models for core-periphery structure.
\newblock In \emph{Proceedings of the twelfth ACM international conference on web search and data mining}, pages 366--374, 2019.

\bibitem[Barber{\'a} et~al.(2015)Barber{\'a}, Wang, Bonneau, Jost, Nagler, Tucker, and Gonz{\'a}lez-Bail{\'o}n]{barbera2015critical}
Pablo Barber{\'a}, Ning Wang, Richard Bonneau, John~T Jost, Jonathan Nagler, Joshua Tucker, and Sandra Gonz{\'a}lez-Bail{\'o}n.
\newblock The critical periphery in the growth of social protests.
\newblock \emph{PloS one}, 10\penalty0 (11):\penalty0 e0143611, 2015.

\bibitem[Bassett et~al.(2013)Bassett, Wymbs, Rombach, Porter, Mucha, and Grafton]{bassett2013task}
Danielle~S Bassett, Nicholas~F Wymbs, M~Puck Rombach, Mason~A Porter, Peter~J Mucha, and Scott~T Grafton.
\newblock Task-based core-periphery organization of human brain dynamics.
\newblock \emph{PLoS computational biology}, 9\penalty0 (9):\penalty0 e1003171, 2013.

\bibitem[Alvarez-Hamelin et~al.(2005)Alvarez-Hamelin, Dall'Asta, Barrat, and Vespignani]{alvarez2005k}
Jos{\'e}~Ignacio Alvarez-Hamelin, Luca Dall'Asta, Alain Barrat, and Alessandro Vespignani.
\newblock K-core decomposition of internet graphs: hierarchies, self-similarity and measurement biases.
\newblock \emph{arXiv preprint cs/0511007}, 2005.

\bibitem[Carmi et~al.(2007)Carmi, Havlin, Kirkpatrick, Shavitt, and Shir]{carmi2007model}
Shai Carmi, Shlomo Havlin, Scott Kirkpatrick, Yuval Shavitt, and Eran Shir.
\newblock A model of internet topology using k-shell decomposition.
\newblock \emph{Proceedings of the National Academy of Sciences}, 104\penalty0 (27):\penalty0 11150--11154, 2007.

\bibitem[Kitsak et~al.(2010)Kitsak, Gallos, Havlin, Liljeros, Muchnik, Stanley, and Makse]{kitsak2010identification}
Maksim Kitsak, Lazaros~K Gallos, Shlomo Havlin, Fredrik Liljeros, Lev Muchnik, H~Eugene Stanley, and Hern{\'a}n~A Makse.
\newblock Identification of influential spreaders in complex networks.
\newblock \emph{Nature physics}, 6\penalty0 (11):\penalty0 888--893, 2010.

\bibitem[Boyd et~al.(2010{\natexlab{b}})Boyd, Fitzgerald, Mahutga, and Smith]{Boyd2010}
John~P Boyd, William~J Fitzgerald, Matthew~C Mahutga, and David~A Smith.
\newblock Computing continuous core/periphery structures for social relations data with minres/svd.
\newblock \emph{Social Networks}, 32\penalty0 (2):\penalty0 125--137, 2010{\natexlab{b}}.

\bibitem[Yang and Leskovec(2014)]{Yang2014overlapping}
Jaewon Yang and Jure Leskovec.
\newblock Overlapping communities explain core--periphery organization of networks.
\newblock \emph{Proceedings of the IEEE}, 102\penalty0 (12):\penalty0 1892--1902, 2014.

\bibitem[Craig and Von~Peter(2010)]{craig2010interbank}
Ben~R. Craig and Goetz Von~Peter.
\newblock Interbank tiering and money center banks.
\newblock 2010.

\bibitem[Van~der Leij et~al.(2016)Van~der Leij, Hommes, and In't~Veld]{van2016formation}
Marco Van~der Leij, Cars~H. Hommes, and Daan In't~Veld.
\newblock The formation of a core-periphery structure in heterogeneous financial networks.
\newblock \emph{De Nederlandsche Bank Working Paper}, \penalty0 (528), 2016.

\bibitem[Ma and Mondrag{\'o}n(2015)]{ma2015rich}
Athen Ma and Ra{\'u}l~J Mondrag{\'o}n.
\newblock Rich-cores in networks.
\newblock \emph{PloS one}, 10\penalty0 (3):\penalty0 e0119678, 2015.

\bibitem[Tun{\c{c}} and Verma(2015{\natexlab{b}})]{Tunc2015}
Birkan Tun{\c{c}} and Ragini Verma.
\newblock Unifying inference of meso-scale structures in networks.
\newblock \emph{PloS one}, 10\penalty0 (11):\penalty0 e0143133, 2015{\natexlab{b}}.

\bibitem[Chang and Zhang(2015)]{chang2015endogenous}
Briana Chang and Shengxing Zhang.
\newblock Endogenous market making and network formation.
\newblock 2015.

\bibitem[Wang(2016)]{wang2016core}
Chaojun Wang.
\newblock Core-periphery trading networks.
\newblock \emph{Available at SSRN 2747117}, 2016.

\bibitem[Bedayo et~al.(2016)Bedayo, Mauleon, and Vannetelbosch]{bedayo2016bargaining}
Mikel Bedayo, Ana Mauleon, and Vincent Vannetelbosch.
\newblock Bargaining in endogenous trading networks.
\newblock \emph{Mathematical Social Sciences}, 80:\penalty0 70--82, 2016.

\bibitem[Castiglionesi and Navarro(2020)]{castiglionesi2020efficient}
Fabio Castiglionesi and Noemi Navarro.
\newblock (in) efficient interbank networks.
\newblock \emph{Journal of Money, Credit and Banking}, 52\penalty0 (2-3):\penalty0 365--407, 2020.

\bibitem[Farboodi(2023)]{farboodi2023intermediation}
Maryam Farboodi.
\newblock Intermediation and voluntary exposure to counterparty risk.
\newblock \emph{Journal of Political Economy}, 131\penalty0 (12):\penalty0 3267--3309, 2023.

\bibitem[Yang(2013)]{yang2013networks}
Song Yang.
\newblock Networks: An introduction by mej newman: Oxford, uk: Oxford university press. 720 pp., 2013.

\bibitem[Boyd et~al.(2006)Boyd, Fitzgerald, and Beck]{boyd2006computing}
John~P Boyd, William~J Fitzgerald, and Robert~J Beck.
\newblock Computing core/periphery structures and permutation tests for social relations data.
\newblock \emph{Social networks}, 28\penalty0 (2):\penalty0 165--178, 2006.

\bibitem[Milo et~al.(2003)Milo, Kashtan, Itzkovitz, Newman, and Alon]{milo2003uniform}
Ron Milo, Nadav Kashtan, Shalev Itzkovitz, Mark~EJ Newman, and Uri Alon.
\newblock On the uniform generation of random graphs with prescribed degree sequences.
\newblock \emph{arXiv preprint cond-mat/0312028}, 2003.

\bibitem[Zlatic et~al.(2009)Zlatic, Bianconi, D{\'\i}az-Guilera, Garlaschelli, Rao, and Caldarelli]{zlatic2009rich}
Vinko Zlatic, Ginestra Bianconi, Albert D{\'\i}az-Guilera, Diego Garlaschelli, Francesco Rao, and Guido Caldarelli.
\newblock On the rich-club effect in dense and weighted networks.
\newblock \emph{The European Physical Journal B}, 67\penalty0 (3):\penalty0 271--275, 2009.

\bibitem[Fortunato(2010)]{fortunato2010community}
Santo Fortunato.
\newblock Community detection in graphs.
\newblock \emph{Physics reports}, 486\penalty0 (3-5):\penalty0 75--174, 2010.

\bibitem[Fortunato and Hric(2016)]{fortunato2016community}
Santo Fortunato and Darko Hric.
\newblock Community detection in networks: A user guide.
\newblock \emph{Physics reports}, 659:\penalty0 1--44, 2016.

\bibitem[Cherifi et~al.(2019)Cherifi, Palla, Szymanski, and Lu]{cherifi2019community}
Hocine Cherifi, Gergely Palla, Boleslaw~K Szymanski, and Xiaoyan Lu.
\newblock On community structure in complex networks: challenges and opportunities.
\newblock \emph{Applied Network Science}, 4\penalty0 (1):\penalty0 1--35, 2019.

\bibitem[Newman(2003{\natexlab{b}})]{newman2003mixing}
Mark~EJ Newman.
\newblock Mixing patterns in networks.
\newblock \emph{Physical review E}, 67\penalty0 (2):\penalty0 026126, 2003{\natexlab{b}}.

\bibitem[Newman(2004{\natexlab{c}})]{newman2004analysis}
Mark~EJ Newman.
\newblock Analysis of weighted networks.
\newblock \emph{Physical Review E—Statistical, Nonlinear, and Soft Matter Physics}, 70\penalty0 (5):\penalty0 056131, 2004{\natexlab{c}}.

\bibitem[Newman(2006{\natexlab{b}})]{newman2006modularity}
Mark~EJ Newman.
\newblock Modularity and community structure in networks.
\newblock \emph{Proceedings of the national academy of sciences}, 103\penalty0 (23):\penalty0 8577--8582, 2006{\natexlab{b}}.

\bibitem[Xu(2020)]{xu2020spectral}
Ying Xu.
\newblock A spectral method to detect community structure based on the communicability modularity.
\newblock \emph{Physica A: Statistical Mechanics and its Applications}, 537:\penalty0 122751, 2020.

\bibitem[Guimera and Nunes~Amaral(2005)]{guimera2005functional}
Roger Guimera and Lu{\'\i}s~A Nunes~Amaral.
\newblock Functional cartography of complex metabolic networks.
\newblock \emph{nature}, 433\penalty0 (7028):\penalty0 895--900, 2005.

\bibitem[Delvenne et~al.(2010)Delvenne, Yaliraki, and Barahona]{delvenne2010stability}
J-C Delvenne, Sophia~N Yaliraki, and Mauricio Barahona.
\newblock Stability of graph communities across time scales.
\newblock \emph{Proceedings of the national academy of sciences}, 107\penalty0 (29):\penalty0 12755--12760, 2010.

\bibitem[Palla et~al.(2005)Palla, Der{\'e}nyi, Farkas, and Vicsek]{palla2005uncovering}
Gergely Palla, Imre Der{\'e}nyi, Ill{\'e}s Farkas, and Tam{\'a}s Vicsek.
\newblock Uncovering the overlapping community structure of complex networks in nature and society.
\newblock \emph{nature}, 435\penalty0 (7043):\penalty0 814--818, 2005.

\bibitem[Zheleva et~al.(2008)Zheleva, Getoor, Golbeck, and Kuter]{zheleva2008using}
Elena Zheleva, Lise Getoor, Jennifer Golbeck, and Ugur Kuter.
\newblock Using friendship ties and family circles for link prediction.
\newblock In \emph{International workshop on social network mining and analysis}, pages 97--113. Springer, 2008.

\bibitem[Lewis et~al.(2010)Lewis, Jones, Porter, and Deane]{lewis2010function}
Anna~CF Lewis, Nick~S Jones, Mason~A Porter, and Charlotte~M Deane.
\newblock The function of communities in protein interaction networks at multiple scales.
\newblock \emph{BMC systems biology}, 4\penalty0 (1):\penalty0 100, 2010.

\bibitem[Zhang et~al.(2010)Zhang, Ning, Ding, and Zhang]{zhang2010determining}
Shihua Zhang, Xue-Mei Ning, Chris Ding, and Xiang-Sun Zhang.
\newblock Determining modular organization of protein interaction networks by maximizing modularity density.
\newblock \emph{BMC systems biology}, 4\penalty0 (Suppl 2):\penalty0 S10, 2010.

\bibitem[Sah et~al.(2014)Sah, Singh, Clauset, and Bansal]{sah2014exploring}
Pratha Sah, Lisa~O Singh, Aaron Clauset, and Shweta Bansal.
\newblock Exploring community structure in biological networks with random graphs.
\newblock \emph{BMC bioinformatics}, 15\penalty0 (1):\penalty0 220, 2014.

\bibitem[Pes{\'a}ntez-Cabrera and Kalyanaraman(2017)]{pesantez2017efficient}
Paola Pes{\'a}ntez-Cabrera and Ananth Kalyanaraman.
\newblock Efficient detection of communities in biological bipartite networks.
\newblock \emph{IEEE/ACM transactions on computational biology and bioinformatics}, 16\penalty0 (1):\penalty0 258--271, 2017.

\bibitem[Rahiminejad et~al.(2019)Rahiminejad, Maurya, and Subramaniam]{rahiminejad2019topological}
Sara Rahiminejad, Mano~R Maurya, and Shankar Subramaniam.
\newblock Topological and functional comparison of community detection algorithms in biological networks.
\newblock \emph{BMC bioinformatics}, 20\penalty0 (1):\penalty0 212, 2019.

\bibitem[Manipur et~al.(2021)Manipur, Giordano, Piccirillo, Parashuraman, and Maddalena]{manipur2021community}
Ichcha Manipur, Maurizio Giordano, Marina Piccirillo, Seetharaman Parashuraman, and Lucia Maddalena.
\newblock Community detection in protein-protein interaction networks and applications.
\newblock \emph{IEEE/ACM Transactions on Computational Biology and Bioinformatics}, 20\penalty0 (1):\penalty0 217--237, 2021.

\bibitem[Harenberg et~al.(2014)Harenberg, Bello, Gjeltema, Ranshous, Harlalka, Seay, Padmanabhan, and Samatova]{harenberg2014community}
Steve Harenberg, Gonzalo Bello, La~Gjeltema, Stephen Ranshous, Jitendra Harlalka, Ramona Seay, Kanchana Padmanabhan, and Nagiza Samatova.
\newblock Community detection in large-scale networks: a survey and empirical evaluation.
\newblock \emph{Wiley Interdisciplinary Reviews: Computational Statistics}, 6\penalty0 (6):\penalty0 426--439, 2014.

\bibitem[Rossetti et~al.(2017)Rossetti, Pappalardo, Pedreschi, and Giannotti]{rossetti2017tiles}
Giulio Rossetti, Luca Pappalardo, Dino Pedreschi, and Fosca Giannotti.
\newblock Tiles: an online algorithm for community discovery in dynamic social networks.
\newblock \emph{Machine Learning}, 106\penalty0 (8):\penalty0 1213--1241, 2017.

\bibitem[Shi et~al.(2017)Shi, Wang, and Lai]{shi2017community}
Wei Shi, Chang-Dong Wang, and Jian-Huang Lai.
\newblock Community detection based on influence power.
\newblock In \emph{Applied Informatics}, volume~4, page~8. Springer, 2017.

\bibitem[Ferretti(2023)]{ferretti2023modeling}
Stefano Ferretti.
\newblock On the modeling and simulation of portfolio allocation schemes: An approach based on network community detection.
\newblock \emph{Computational Economics}, 62\penalty0 (3):\penalty0 969--1005, 2023.

\bibitem[Yang et~al.(2023)Yang, Liu, Li, and Xie]{yang2023communities}
Xuenan Yang, Zihan Liu, Jingyu Li, and Qiwei Xie.
\newblock Communities of co-occurrence network of financial firms in news.
\newblock \emph{Procedia Computer Science}, 221:\penalty0 821--825, 2023.

\bibitem[Fenn et~al.(2009)Fenn, Porter, McDonald, Williams, Johnson, and Jones]{fenn2009dynamic}
Daniel~J Fenn, Mason~A Porter, Mark McDonald, Stacy Williams, Neil~F Johnson, and Nick~S Jones.
\newblock Dynamic communities in multichannel data: An application to the foreign exchange market during the 2007--2008 credit crisis.
\newblock \emph{CHAOS}, 19:\penalty0 033119, 2009.

\bibitem[Barucca and Lillo(2016)]{barucca2016disentangling}
Paolo Barucca and Fabrizio Lillo.
\newblock Disentangling bipartite and core-periphery structure in financial networks.
\newblock \emph{Chaos, Solitons \& Fractals}, 88:\penalty0 244--253, 2016.
\newblock \doi{10.1016/j.chaos.2016.03.013}.

\bibitem[Guimerà and Amaral(2005)]{guimera2005worldwide}
Roger Guimerà and Luís A.~N. Amaral.
\newblock The worldwide air transportation network: Anomalous centrality, community structure, and cities' global roles.
\newblock \emph{Proceedings of the National Academy of Sciences}, 102\penalty0 (22):\penalty0 7794--7799, 2005.
\newblock \doi{10.1073/pnas.0407994102}.

\bibitem[Derrible(2012)]{derrible2012network}
Sybil Derrible.
\newblock Network centrality of metro systems.
\newblock \emph{PLoS ONE}, 7\penalty0 (7):\penalty0 e40575, 2012.
\newblock \doi{10.1371/journal.pone.0040575}.

\bibitem[Hu et~al.(2010)Hu, Nie, Yang, Cheng, Fan, and Di]{hu2010measuring}
Yanqing Hu, Yuchao Nie, Hua Yang, Jie Cheng, Ying Fan, and Zengru Di.
\newblock Measuring the significance of community structure in complex networks.
\newblock \emph{Physical Review E—Statistical, Nonlinear, and Soft Matter Physics}, 82\penalty0 (6):\penalty0 066106, 2010.

\bibitem[Lancichinetti et~al.(2010)Lancichinetti, Radicchi, and Ramasco]{lancichinetti2010statistical}
Andrea Lancichinetti, Filippo Radicchi, and Jos{\'e}~J Ramasco.
\newblock Statistical significance of communities in networks.
\newblock \emph{Physical Review E—Statistical, Nonlinear, and Soft Matter Physics}, 81\penalty0 (4):\penalty0 046110, 2010.

\bibitem[Lancichinetti et~al.(2011)Lancichinetti, Radicchi, Ramasco, and Fortunato]{lancichinetti2011finding}
Andrea Lancichinetti, Filippo Radicchi, Jos{\'e}~J Ramasco, and Santo Fortunato.
\newblock Finding statistically significant communities in networks.
\newblock \emph{PloS one}, 6\penalty0 (4):\penalty0 e18961, 2011.

\bibitem[Zhang and Chen(2017)]{zhang2017hypothesis}
Jingfei Zhang and Yuguo Chen.
\newblock A hypothesis testing framework for modularity based network community detection.
\newblock \emph{Statistica Sinica}, pages 437--456, 2017.

\bibitem[Kojaku and Masuda(2018{\natexlab{b}})]{kojaku2018generalised}
Sadamori Kojaku and Naoki Masuda.
\newblock A generalised significance test for individual communities in networks.
\newblock \emph{Scientific reports}, 8\penalty0 (1):\penalty0 7351, 2018{\natexlab{b}}.

\bibitem[Palowitch et~al.(2018)Palowitch, Bhamidi, and Nobel]{palowitch2018significance}
John Palowitch, Shankar Bhamidi, and Andrew~B Nobel.
\newblock Significance-based community detection in weighted networks.
\newblock \emph{Journal of Machine Learning Research}, 18\penalty0 (188):\penalty0 1--48, 2018.

\bibitem[Yanchenko and Sengupta(2024)]{yanchenko2024generalized}
Eric Yanchenko and Srijan Sengupta.
\newblock A generalized hypothesis test for community structure in networks.
\newblock \emph{Network Science}, 12\penalty0 (2):\penalty0 122--138, 2024.

\bibitem[Ansari et~al.(2025{\natexlab{a}})Ansari, Sharma, Agrawal, and Sahni]{ansari2025novel}
Imran Ansari, Charu Sharma, Akshay Agrawal, and Niteesh Sahni.
\newblock A novel portfolio construction strategy based on the core-periphery profile of stocks.
\newblock \emph{Scientific Reports}, 15\penalty0 (1):\penalty0 1--28, 2025{\natexlab{a}}.

\bibitem[Pozzi et~al.(2013)Pozzi, Di~Matteo, and Aste]{pozzi2013spread}
Francesco Pozzi, Tiziana Di~Matteo, and Tomaso Aste.
\newblock Spread of risk across financial markets: better to invest in the peripheries.
\newblock \emph{Scientific reports}, 3\penalty0 (1):\penalty0 1665, 2013.

\bibitem[Li et~al.(2019)Li, Jiang, Tian, Li, and Zheng]{li2019portfolio}
Yan Li, Xiong-Fei Jiang, Yue Tian, Sai-Ping Li, and Bo~Zheng.
\newblock Portfolio optimization based on network topology.
\newblock \emph{Physica A: Statistical Mechanics and its Applications}, 515:\penalty0 671--681, 2019.

\bibitem[Sharma and Habib(2019)]{sharma2019mutual}
Charu Sharma and Amber Habib.
\newblock Mutual information based stock networks and portfolio selection for intraday traders using high frequency data: An indian market case study.
\newblock \emph{PloS one}, 14\penalty0 (8):\penalty0 e0221910, 2019.

\bibitem[Rosvall and Bergstrom(2011)]{rosvall2011multilevel}
Martin Rosvall and Carl~T Bergstrom.
\newblock Multilevel compression of random walks on networks reveals hierarchical organization in large integrated systems.
\newblock \emph{PloS one}, 6\penalty0 (4):\penalty0 e18209, 2011.

\bibitem[Tumminello et~al.(2005)Tumminello, Aste, Di~Matteo, and Mantegna]{tumminello2005tool}
Michele Tumminello, Tomaso Aste, Tiziana Di~Matteo, and Rosario~N Mantegna.
\newblock A tool for filtering information in complex systems.
\newblock \emph{Proceedings of the National Academy of Sciences}, 102\penalty0 (30):\penalty0 10421--10426, 2005.

\bibitem[Pozzi et~al.(2012)Pozzi, Di~Matteo, and Aste]{pozzi2012exponential}
Francesco Pozzi, Tiziana Di~Matteo, and Tomaso Aste.
\newblock Exponential smoothing weighted correlations.
\newblock \emph{The European Physical Journal B}, 85:\penalty0 1--21, 2012.

\bibitem[Milgram(1967)]{milgram1967small}
Stanley Milgram.
\newblock The small world problem.
\newblock \emph{Psychology today}, 2\penalty0 (1):\penalty0 60--67, 1967.

\bibitem[Faloutsos et~al.(1999)Faloutsos, Faloutsos, and Faloutsos]{faloutsos1999power}
Michalis Faloutsos, Petros Faloutsos, and Christos Faloutsos.
\newblock On power-law relationships of the internet topology.
\newblock \emph{ACM SIGCOMM computer communication review}, 29\penalty0 (4):\penalty0 251--262, 1999.

\bibitem[Papadopoulos et~al.(2012)Papadopoulos, Kitsak, Serrano, Bogun{\'a}, and Krioukov]{papadopoulos2012popularity}
Fragkiskos Papadopoulos, Maksim Kitsak, M~{\'A}ngeles Serrano, Mari{\'a}n Bogun{\'a}, and Dmitri Krioukov.
\newblock Popularity versus similarity in growing networks.
\newblock \emph{Nature}, 489\penalty0 (7417):\penalty0 537--540, 2012.

\bibitem[Serrano et~al.(2008)Serrano, Krioukov, and Bogun{\'a}]{serrano2008self}
M~{\'A}ngeles Serrano, Dmitri Krioukov, and Mari{\'a}n Bogun{\'a}.
\newblock Self-similarity of complex networks and hidden metric spaces.
\newblock \emph{Physical review letters}, 100\penalty0 (7):\penalty0 078701, 2008.

\bibitem[Garc{\'i}a-P{\'e}rez et~al.(2019)Garc{\'i}a-P{\'e}rez, Allard, Serrano, and Bogu{\~n}{\'a}]{garcia2019mercator}
Guillermo Garc{\'i}a-P{\'e}rez, Antoine Allard, Mari{\'a}n~{\'A}. Serrano, and Mari{\'a}n Bogu{\~n}{\'a}.
\newblock Mercator: uncovering faithful hyperbolic embeddings of complex networks.
\newblock \emph{New Journal of Physics}, 21\penalty0 (12):\penalty0 123033, 2019.

\bibitem[Ansari et~al.(2025{\natexlab{b}})Ansari, Pawanesh, and Sahni]{ansari2025uncovering}
Imran Ansari, Pawanesh Pawanesh, and Niteesh Sahni.
\newblock Uncovering the hidden core-periphery structure in hyperbolic networks.
\newblock \emph{Physical Review E}, 112\penalty0 (3):\penalty0 034311, 2025{\natexlab{b}}.

\bibitem[Papadopoulos et~al.(2014)Papadopoulos, Psomas, and Krioukov]{papadopoulos2014network}
Fragkiskos Papadopoulos, Constantinos Psomas, and Dmitri Krioukov.
\newblock Network mapping by replaying hyperbolic growth.
\newblock \emph{IEEE/ACM Transactions on Networking}, 23\penalty0 (1):\penalty0 198--211, 2014.

\bibitem[Wedell et~al.(2022)Wedell, Park, Korobskiy, Warnow, and Chacko]{wedell2022center}
Eleanor Wedell, Minhyuk Park, Dmitriy Korobskiy, Tandy Warnow, and George Chacko.
\newblock Center--periphery structure in research communities.
\newblock \emph{Quantitative Science Studies}, 3\penalty0 (1):\penalty0 289--314, 2022.

\bibitem[Peixoto and Rosvall(2017)]{peixoto2017modelling}
Tiago~P Peixoto and Martin Rosvall.
\newblock Modelling sequences and temporal networks with dynamic community structures.
\newblock \emph{Nature communications}, 8\penalty0 (1):\penalty0 582, 2017.

\bibitem[Gauvin et~al.(2014)Gauvin, Panisson, and Cattuto]{gauvin2014detecting}
Laetitia Gauvin, Andr{\'e} Panisson, and Ciro Cattuto.
\newblock Detecting the community structure and activity patterns of temporal networks: a non-negative tensor factorization approach.
\newblock \emph{PloS one}, 9\penalty0 (1):\penalty0 e86028, 2014.

\bibitem[He and Chen(2015)]{he2015fast}
Jialin He and Duanbing Chen.
\newblock A fast algorithm for community detection in temporal network.
\newblock \emph{Physica A: Statistical Mechanics and its Applications}, 429:\penalty0 87--94, 2015.

\bibitem[Linhares et~al.(2019)Linhares, Ponciano, Pereira, Rocha, Paiva, and Traven{\c{c}}olo]{linhares2019scalable}
Claudio~DG Linhares, Jean~R Ponciano, Fabiola~SF Pereira, Luis~EC Rocha, Jose Gustavo~S Paiva, and Bruno~AN Traven{\c{c}}olo.
\newblock A scalable node ordering strategy based on community structure for enhanced temporal network visualization.
\newblock \emph{Computers \& Graphics}, 84:\penalty0 185--198, 2019.

\bibitem[Jeub et~al.(2017)Jeub, Mahoney, Mucha, and Porter]{jeub2017local}
Lucas~GS Jeub, Michael~W Mahoney, Peter~J Mucha, and Mason~A Porter.
\newblock A local perspective on community structure in multilayer networks.
\newblock \emph{Network Science}, 5\penalty0 (2):\penalty0 144--163, 2017.

\bibitem[Pramanik et~al.(2017)Pramanik, Tackx, Navelkar, Guillaume, and Mitra]{pramanik2017discovering}
Soumajit Pramanik, Raphael Tackx, Anchit Navelkar, Jean-Loup Guillaume, and Bivas Mitra.
\newblock Discovering community structure in multilayer networks.
\newblock In \emph{2017 IEEE international conference on data science and advanced analytics (DSAA)}, pages 611--620. IEEE, 2017.

\bibitem[Khawaja et~al.(2021)Khawaja, Sheng, Wang, and Memon]{khawaja2021uncovering}
Faiza~Riaz Khawaja, Jinfang Sheng, Bin Wang, and Yumna Memon.
\newblock Uncovering hidden community structure in multi-layer networks.
\newblock \emph{Applied Sciences}, 11\penalty0 (6):\penalty0 2857, 2021.

\bibitem[Wilson et~al.(2017)Wilson, Palowitch, Bhamidi, and Nobel]{wilson2017community}
James~D Wilson, John Palowitch, Shankar Bhamidi, and Andrew~B Nobel.
\newblock Community extraction in multilayer networks with heterogeneous community structure.
\newblock \emph{Journal of Machine Learning Research}, 18\penalty0 (149):\penalty0 1--49, 2017.

\bibitem[Bergermann and Tudisco(2025)]{bergermann2025core}
Kai Bergermann and Francesco Tudisco.
\newblock Core-periphery detection in multilayer networks.
\newblock \emph{Physical Review Letters}, 135\penalty0 (8):\penalty0 087402, 2025.

\bibitem[Beranek and Remes(2023)]{beranek2023emergence}
L~Beranek and R~Remes.
\newblock The emergence of a core--periphery structure in evolving multilayer network.
\newblock \emph{Physica A: Statistical Mechanics and its Applications}, 612:\penalty0 128484, 2023.

\bibitem[Pan and Sinha(2007)]{pan2007collective}
Raj~Kumar Pan and Sitabhra Sinha.
\newblock Collective behavior of stock price movements in an emerging market.
\newblock \emph{Physical Review E—Statistical, Nonlinear, and Soft Matter Physics}, 76\penalty0 (4):\penalty0 046116, 2007.

\bibitem[Jiang et~al.(2014)Jiang, Chen, and Zheng]{jiang2014structure}
Xiongfei Jiang, Tianshou Chen, and Baowen Zheng.
\newblock Structure of local interactions in complex financial dynamics.
\newblock \emph{Scientific Reports}, 4:\penalty0 5321, 2014.

\bibitem[Kov{\'a}cs et~al.(2022)Kov{\'a}cs, Balogh, and Palla]{kovacs2022generalised}
B{\'a}lint Kov{\'a}cs, S{\'a}ndor~G. Balogh, and Gergely Palla.
\newblock Generalised popularity-similarity optimisation model for growing hyperbolic networks beyond two dimensions.
\newblock \emph{Scientific Reports}, 12\penalty0 (1):\penalty0 968, 2022.

\bibitem[Jankowski et~al.(2023)Jankowski, Allard, Bogu{\~n}{\'a}, and Serrano]{jankowski2023d}
Robert Jankowski, Antoine Allard, Mari{\'a}n Bogu{\~n}{\'a}, and M{\'a}rio~{\'A}. Serrano.
\newblock The d-mercator method for the multidimensional hyperbolic embedding of real networks.
\newblock \emph{Nature Communications}, 14\penalty0 (1):\penalty0 7585, 2023.

\bibitem[Budel et~al.(2024)Budel, Kitsak, Aldecoa, Zuev, and Krioukov]{budel2024random}
Gilad Budel, Maksim Kitsak, Rodrigo Aldecoa, Konstantin Zuev, and Dmitri Krioukov.
\newblock Random hyperbolic graphs in d+ 1 dimensions.
\newblock \emph{Physical Review E}, 109\penalty0 (5):\penalty0 054131, 2024.

\bibitem[Tudisco and Higham(2023)]{tudisco2023core}
Francesco Tudisco and Desmond~J Higham.
\newblock Core-periphery detection in hypergraphs.
\newblock \emph{SIAM Journal on Mathematics of Data Science}, 5\penalty0 (1):\penalty0 1--21, 2023.

\bibitem[Papachristou and Kleinberg(2022)]{papachristou2022core}
Marios Papachristou and Jon Kleinberg.
\newblock Core-periphery models for hypergraphs.
\newblock In \emph{Proceedings of the 28th ACM SIGKDD Conference on Knowledge Discovery and Data Mining}, pages 1337--1347, 2022.

\bibitem[Yuan et~al.(2022)Yuan, Liu, Feng, and Shang]{yuan2022testing}
Mingao Yuan, Ruiqi Liu, Yang Feng, and Zuofeng Shang.
\newblock Testing community structure for hypergraphs.
\newblock \emph{The Annals of Statistics}, 50\penalty0 (1):\penalty0 147--169, 2022.

\bibitem[Ruggeri et~al.(2024)Ruggeri, Battiston, and De~Bacco]{ruggeri2024framework}
Nicol{\`o} Ruggeri, Federico Battiston, and Caterina De~Bacco.
\newblock Framework to generate hypergraphs with community structure.
\newblock \emph{Physical Review E}, 109\penalty0 (3):\penalty0 034309, 2024.

\end{thebibliography}


\end{document}